\documentclass[preprint,12pt]{elsarticle}

\usepackage{amssymb}
\usepackage[export]{adjustbox}

\usepackage[%
    colorlinks=true,
    pdfborder={0 0 0},
    linkcolor=red
]{hyperref}
 \usepackage{amsthm}
 
\usepackage{float}
\usepackage{amsmath}
\usepackage{array}
\usepackage{booktabs}
\usepackage{subcaption}
\usepackage{siunitx}
\usepackage{graphicx}
\newcommand{\fluenceUnits}[1]{\SI{#1}{n_{eq}\,\cm^{-2}}}
\newcommand{\lumiUnits}[1]{\SI{#1}{\cm^{-2}\,\second^{-1}}}

\journal{NIM-A}

\begin{document}

\begin{frontmatter}



\title{Effects of Neutron Irradiation on LGADs with Broad Multiplication Layer and varied Carbon-Enriched Doses: A Study on Timing Performance and Gain Deterioration}

\author[a]{E. Navarrete Ramos \footnote{Corresponding author: efren.navarrete@unican.es}}
\author[b]{J. Villegas}
\author[a]{J. Duarte-Campderros}
\author[a]{M. Fernández}
\author[a]{A. Gómez-Carrera}
\author[a]{G. Gómez}
\author[a]{J. González}
\author[c]{S. Hidalgo}
\author[a]{R. Jaramillo}
\author[a]{P. Martínez Ruiz del Árbol}
\author[c]{A. Merlos}
\author[a]{C. Quintana}
\author[a]{I. Vila}


\affiliation[a]{organization={Instituto de Física de Cantabria, IFCA (CSIC-UC)},
            addressline={Av. los Castros},
            city={Santander},
            postcode={39005},
            country={Spain}}

\affiliation[b]{organization={Centro Nacional de Aceleradores, CNA},
             city={Seville},
             postcode={41092},
             country={Spain}}

\affiliation[c]{organization={Instituto de Microelectrónica de Barcelona, IMB-CNM (CSIC)},
             addressline={C/ dels Til·lers Cerdanyola del Vallès},
             city={Barcelona},
             postcode={08193},
             country={Spain}}


\begin{abstract}
    
In this radiation tolerance study, Low Gain Avalanche Detectors (LGADs) with a carbon-enriched broad and shallow multiplication layer were examined in comparison to identical non-carbonated LGADs. Manufactured at IMB-CNM, the sensors underwent neutron irradiation at the TRIGRA reactor in Ljubljana, reaching a fluence of \fluenceUnits{2.5e15}. The results revealed a smaller deactivation of Boron and improved resistance to radiation in carbonated LGADs. The study demonstrated the potential benefits of carbon enrichment in mitigating radiation damage effects, particularly the acceptor removal mechanism, reducing the acceptor removal constant by more than a factor of two. Additionally, time resolution and collected charge degradation due to irradiation were observed, with carbonated samples exhibiting better radiation tolerance. A noise analysis focused on baseline noise and thermally generated pulses showed the presence of spurious thermal-generated pulses attributed to a excessive small distance between the gain layer end and the p-stop implant at the periphery of the pad for the characterized LGAD design; however, the operation performance of the devices was unaffected.

\end{abstract}


\end{frontmatter}


\section{Introduction}
\label{sec:intro}

The Large Hadron Collider (HL-LHC) is set to undergo a high-luminosity upgrade, which is planned to commence in early 2029. This upgrade is projected to deliver an integrated luminosity of up to \SI{4000}{\per\femto\barn} over a decade~\cite{Aberle:2749422}. The HL-LHC is designed to operate at a stable luminosity of \lumiUnits{5.0e34}, with a potential maximum of \lumiUnits{7.5e{34}}. One of the primary challenges of the HL-LHC will be managing the superposition of multiple proton-proton collisions per bunch crossing, referred to as \emph{pileup}, within a confined region. This region of multiple collisions is expected to extend approximately \SI{50}{\milli\meter} RMS along the beam axis, with an average of \SI{1.6}{collisions\per\milli\meter} and up to \SI{200}{pp} interactions per bunch crossing. 

In such conditions, separating the multiple collisions and accurately associating the reconstructed tracks to their originating vertex will pose a significant challenge. To tackle this, MIP timing sub-detectors have been proposed~\cite{CERN-LHCC-2017-027, CERN-LHCC-2018-023}, which are anticipated to offer a time resolution of \SI{30}{\pico\second} per track. These detectors are predicted to considerably enhance the performance of the ATLAS and CMS detectors by disentangling the high number of pileup events.

The CMS Endcap Timing layer (ETL), a proposed sub-detector, is planned to be constructed using Low Gain Avalanche Detectors (LGAD) with a pixel size of \qtyproduct{1.3 x 1.3}{\milli\meter\squared}. The ETL is designed to cover the pseudorapidity range of $1.6<|\eta|<3.0$, with a total surface area of \SI{14}{\meter\squared}. This sub-detector will be subjected to radiation levels up to \fluenceUnits{1.5e15} at $|\eta|=3.0$. However, for \SI{80}{\percent} of the ETL area, the fluence is expected to be less than \fluenceUnits{1e15}. Hence, these two fluence points are the focus of this radiation tolerance study.

\begin{figure}
     \centering
         \includegraphics[width=0.8\textwidth]{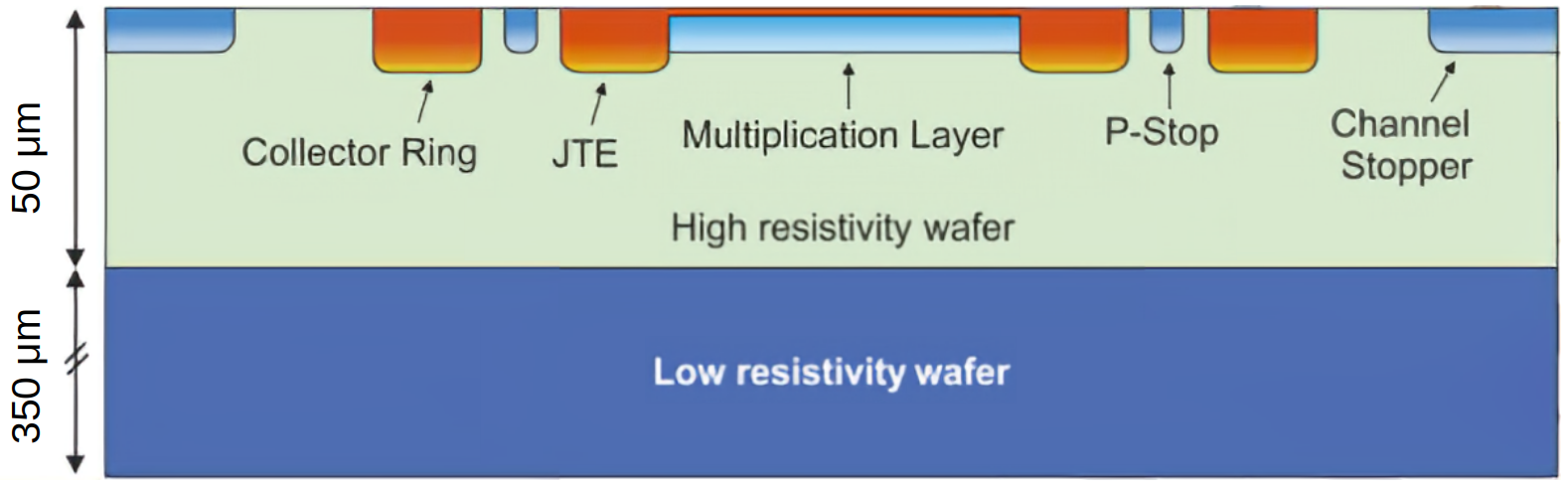}
       \caption{Description of the cross-sectional structure of a Low Gain Avalanche Detector. Components like the  Collector Ring, Channel Stopper, P-Stop, Multiplication Layer, and the Junction Termination Extension (JTE) are illustrated. Note that the thickness of the active volume, low resistivity wafer, and component distributions are not scaled proportionally.}
        \label{fig:scheme}
\end{figure}

LGADs are semiconductor detectors engineered for signal amplification commonly built as $n^{++}-p^+-p$ avalanche diodes, with a highly-doped $p^{+}$ layer introduced to create a region of extremely high electric field. This electric field initiates the avalanche multiplication of primary electrons, producing additional electron-hole pairs. Cross-section of a standard pad-like LGAD is depicted in \autoref{fig:scheme}. The structure of the LGAD is meticulously designed to achieve a moderate gain and function effectively across a broad range of reverse bias voltages before the so called breakdown regime.

This document presents a radiation tolerance study conducted on LGAD devices with carbon-enriched multiplication layer. Its performance was evaluated against LGADs with the same layout and manufacturing process, but with different carbon enrichment dose. The LGAD sensors were fabricated at IMB-CNM (Institute of Microelectronics of Barcelona, Spain)~\cite{CNM} and irradiated with neutrons in Ljubljana with the TRIGRA reactor up to a fluence of \fluenceUnits{2.5e15}. This study reports on the degradation of its time resolution performance and charge collection due to irradiation.

\section{Samples Description}
\label{sec:descript}

This work is focused on the LGAD production of the IMB-CNM based in different devices with different carbon enrichment doses on its broad multiplication layer in order to study their radiation tolerance. The LGADs from Run\#15973 were produced on Si-Si (silicon on silicon) wafers with a diameter of 6 inches. The active layer thickness is \SI{50}{\micro\meter}, and the handle wafer is \SI{350}{\micro\meter} thick. The resistivity \SIrange{0.001}{1}{\ohm\centi\meter}, while the substrate resistivity is approximately \SI{2}{k\ohm\cm}. This run is the second incorporating carbon enrichment of the gain layer in some wafers (Run\#15246~\cite{Navarrete15246} was the first). These wafers include matrices with varying numbers of pads: 1$\times$1 (single pad diodes), 2$\times$2, 5$\times$5, and 15$\times$15, with each pad of \qtyproduct{1.3 x 1.3}{\mm\squared} of area.

The manufacturing parameters of the wafers, including the Boron and the different Carbon doses, are detailed in \autoref{tab:wafers}. It’s worth noting that the main distinction between these samples is the carbon dose to the gain layer and from here on we will refer to them as Low Carbonated (with a carbon dose of (\SI{3e14}{\per\cm\squared}) and High Carbonated (\SI{9e14}{\per\cm\squared}) samples depending on this parameter, as well as Standard to the non-carbonated samples. The use of carbon co-implantation in the gain layer during the manufacturing of LGAD sensors effectively mitigates the effects of the acceptor removal mechanism~\cite{Ferrero2019}, which signifies the degradation of the gain due to radiation damage. The results on the acceptor removal effect on these carbonated sensors are presented in \autoref{sec:acceptor}.

\begin{table}[htbp]
\centering
\caption{\label{tab:wafers} Differences in depletion voltages and carbon and boron doses for the different devices}
\smallskip
\begin{tabular}{m{13em} m{1.5cm} m{1.5cm} m{1.5cm} }
\hline
Wafer & Standard & Low C &  High C \\
\hline\hline

Boron energy (keV)   & 100 & 100 & 100 \\

Boron dose (\SI{1e13}{\per\cm\squared})& 1.9 & 1.9 & 1.9\\

Carbon energy (keV)  & 150 & 150 & 150 \\

Carbon dose (\SI{1e14}{\per\cm\squared})& 0 & 3 & 9 \\
\hline
\end{tabular}
\end{table}

A characterization campaign were carried out at the Instituto de Física de Cantabria (IFCA) to evaluate sensors with different carbon doses. A summary of the sensors measured in the radioactive source setup is provided in \autoref{tab:rs-sensors}. For the radioactive source measurements, the samples were organized in a stack of three sensors, with one non-irradiated sensor serving as a time reference. Some samples were irradiated with neutrons to four different fluences: \fluenceUnits{0.4e15}, \fluenceUnits{0.8e15}, \fluenceUnits{1.5e15} and \fluenceUnits{2.5e15}, in the \SI{250}{\kilo\watt} TRIGA Mark II reactor\footnote{which has a maximum flux of around \SI{2e13}{n\,\cm^{-2}\,\second^{-1}}~\cite{TRIGA} at its center.} of the Jožef Stefan Institute (JSI)~\cite{Zontar}  at Ljubljana Slovenia. The number of sensors measured in electrical characterization is bigger than the sensors measured in the RS setup. In summary, Two (Three) samples were measured in the beta source setup at IFCA after (before) Irradiation.

\begin{table}[htbp]
\centering
\caption{\label{tab:rs-sensors} LGAD samples of each fluence point characterized at IFCA}
\smallskip
\begin{tabular}{ l c c c }
\hline
Fluence (\fluenceUnits{ }) & Standard & Low C & High C  \\
\hline\hline
0 & 3 & 3 & 3 \\

$ 4\times 10^{14} $ & 2 & 2 & 2 \\

$ 8\times 10^{14} $ & 2 & 2 & 2 \\

$ 15\times 10^{14}$ & 2 & 2 & 2 \\

$ 25\times 10^{14}$ & 2 & 2 & 2  \\
\hline
\end{tabular}
\end{table}

\section{Electrical Characterization}
\label{sec:Electric}

The characteristics of Current-Voltage (IV) and Capacitance-Voltage (CV) were assessed in a probe station outfitted with a thermal chuck. These assessments were conducted both pre and post irradiation. Non-irradiated devices were measured at room temperature, whereas the irradiated devices were evaluated at a temperature of \SI{-25}{\celsius}. The backside (ohmic contact) of the sensors was grounded, and the GR and cathode were connected to High-Voltage (HV). For the IV assessment, the currents of the main pad and the GR were independently determined using two distinct Keithley 2410 sourcemeters~\cite{keith}, which facilitated the supply of High-Voltage to the device and simultaneous current measurement. For the CV assessment, the GR and the main diode were HV biased using Keithley 2410 sourcemeters and read by a Quadtech 1920 LCR-meter~\cite{LCR} via a decoupling box. The capacitance was determined using a parallel RC circuit model, and the measurements were performed at \SI{100}{\hertz} frequency before and after irradiation.

The samples are represented by colors, with black denoting the standard sensors (without carbon enrichment: \SI{0e14}{\per\cm\squared}), blue for the Low Carbonated  sensors with a carbon enrichment of \SI{3e14}{\per\cm\squared} and red for the High Carbonated sensors with \SI{9e14}{\per\cm\squared}. This color coding is delineated in the accompanying legends. The fluence of the irradiation, if applicable, is specified in the footer of each figure.

\subsection{Current-Voltage characteristic}
\label{sec:iv}

\begin{figure}
     \centering
     \begin{subfigure}[b]{0.49\textwidth}
         \centering
         \includegraphics[width=\textwidth, trim={0.6cm 0.4cm 1.5cm 0.75cm}, clip=true]{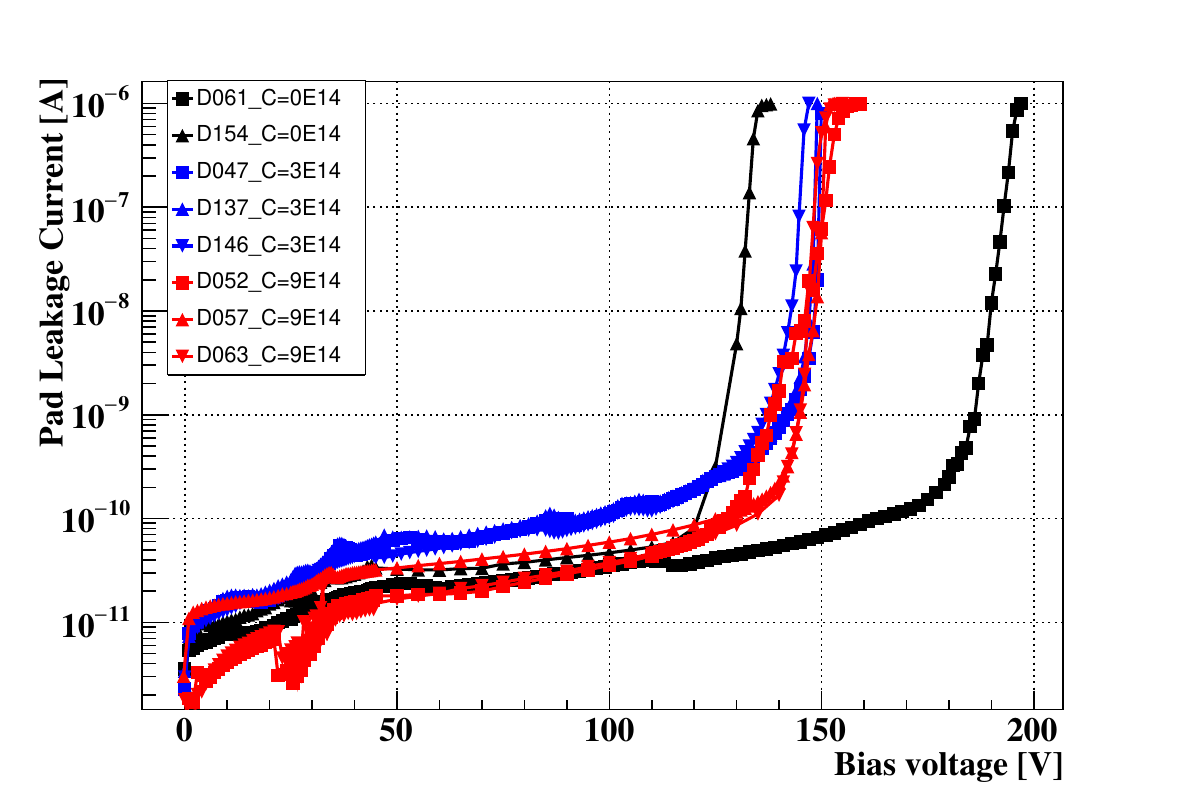}
         \caption{IV curves}
     \end{subfigure}
     \hfill
     \begin{subfigure}[b]{0.49\textwidth}
         \centering
         \includegraphics[width=\textwidth, trim={0.6cm 0.4cm 1.5cm 0.75cm}, clip=true]{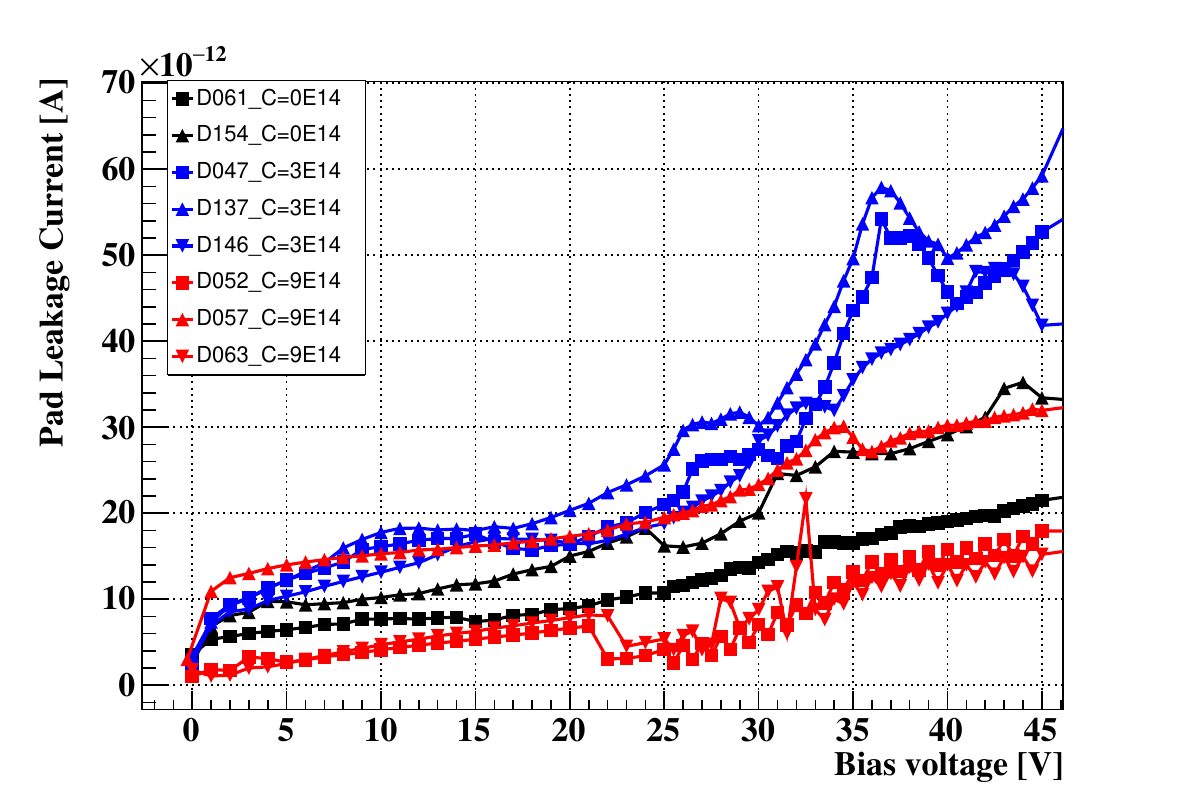}
         \caption{IV curves GL region}
     \end{subfigure}
         \caption{The leakage currents of the main diode before irradiation are presented as a function of reverse bias before irradiation. The left-hand plot (a) shows the complete IV curve, while the right-hand plots (b) it is an enlarged view of the region where the gain layer is depleted.}
        \label{fig:IV_pre}
\end{figure}

The main diode leakage current versus the bias voltage for the different type of sensors prior to irradiation and at room temperature is illustrated in \autoref{fig:IV_pre}. The current of all the samples is below the nanoampere across the majority of the bias voltage range before the breakdown. The breakdown voltage $V_{BD}$ was calculated by estimating the change in the slope using the method outlined in \autoref{sec:acceptor}, and it was found to be between \SI{130}{\volt} and \SI{150}{\volt}, except for one of the non-irradiated samples that has an unexpected larger $V_{BD}$ of about \SI{190}{\volt}. In general, the $V_{BD}$ for the three wafers studied has a low dispersion and as expected, the leakage current of the carbonated sensors in the gain layer region is higher than that of standard sensors, since the carbon enhancement increases the defects in the gain layer~\cite{Vag_gl}.

\begin{figure}
     \centering
     \begin{subfigure}[b]{0.49\textwidth}
         \centering
         \includegraphics[width=\textwidth, trim={0.6cm 0.4cm 1.5cm 0.7cm}, clip=true]{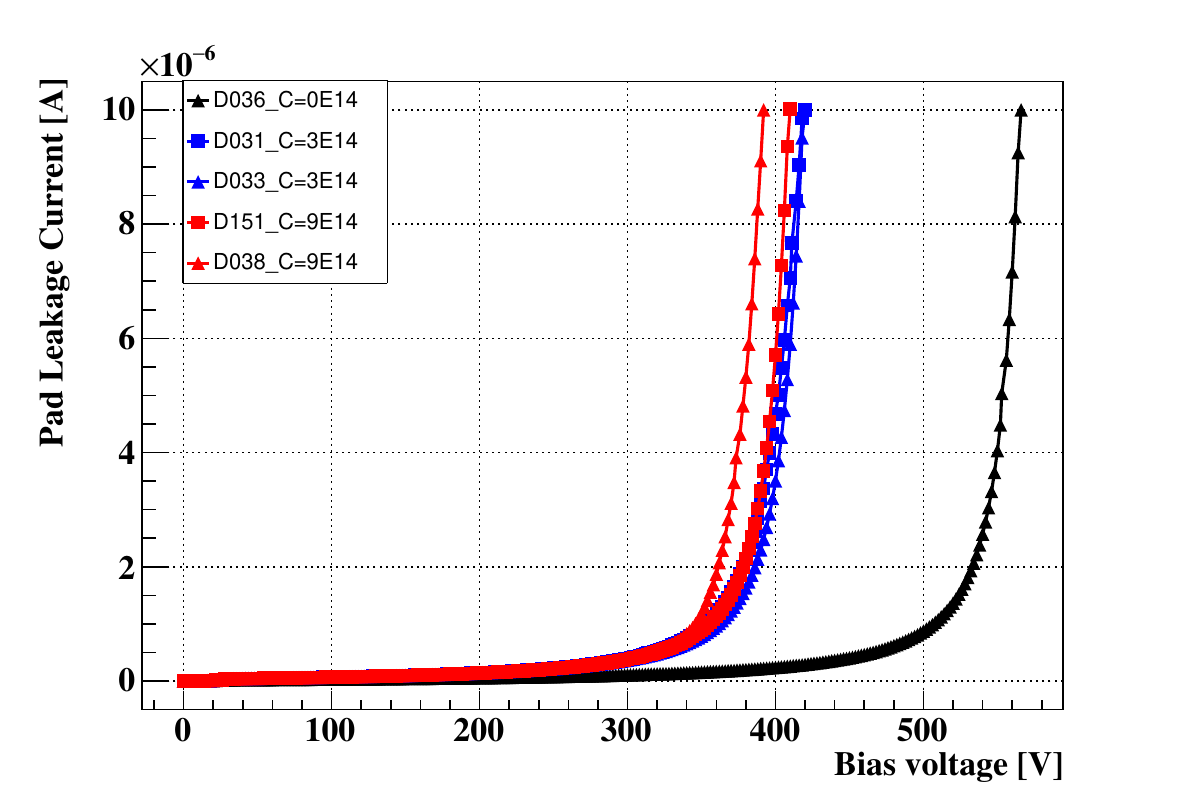}
         \caption{Samples irradiated at \fluenceUnits{4e14}}
     \end{subfigure}
     \hfill
     \begin{subfigure}[b]{0.49\textwidth}
         \centering
         \includegraphics[width=\textwidth, trim={0.6cm 0.4cm 1.5cm 0.7cm}, clip=true]{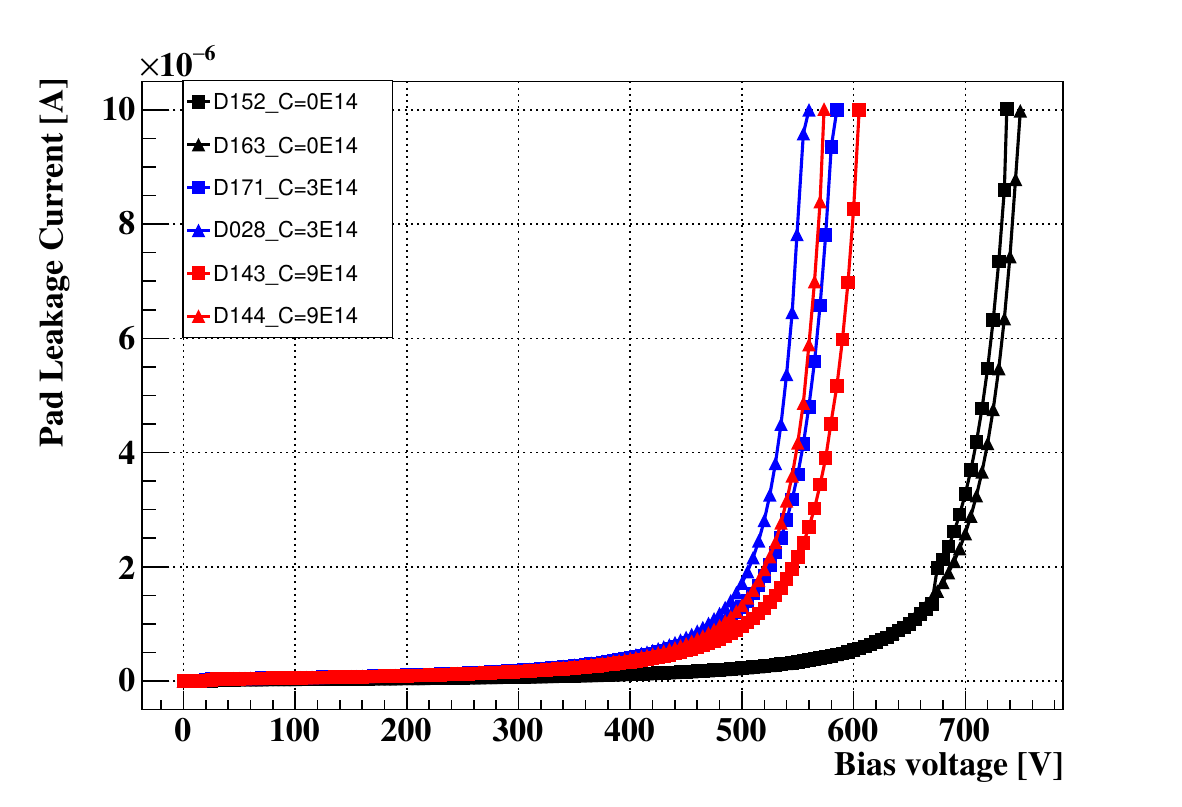}
         \caption{Samples irradiated at \fluenceUnits{8e14}}
     \end{subfigure}
     \begin{subfigure}[b]{0.49\textwidth}
         \centering
         \includegraphics[width=\textwidth, trim={0.6cm 0.4cm 1.5cm 0.7cm}, clip=true]{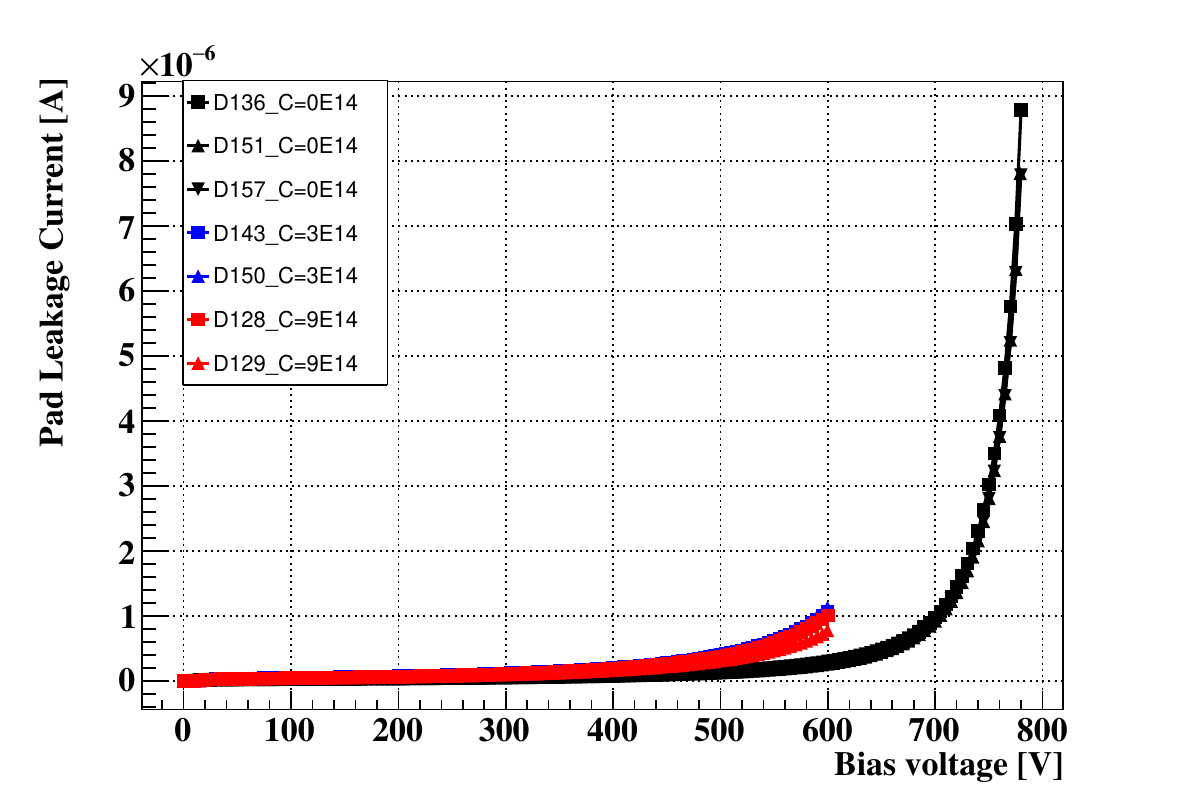}
         \caption{Samples irradiated at \fluenceUnits{15e14}}
     \end{subfigure}
     \hfill
     \begin{subfigure}[b]{0.49\textwidth}
         \centering
         \includegraphics[width=\textwidth, trim={0.6cm 0.4cm 1.5cm 0.7cm}, clip=true]{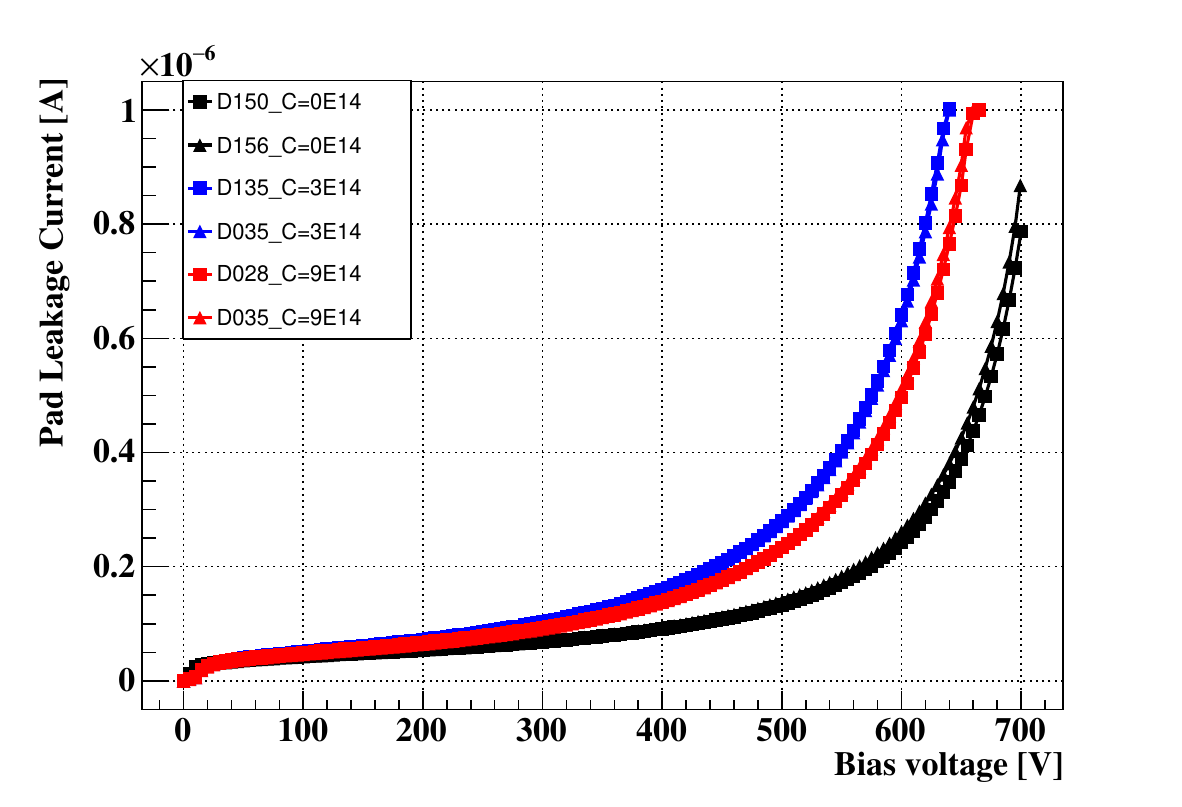}
         \caption{Samples irradiated at \fluenceUnits{25e14}}
     \end{subfigure}

        \caption{The leakage currents of the main pad after irradiation as a function of the reverse bias is shown. Samples irradiated at \fluenceUnits{4e14}, \fluenceUnits{8e14}, \fluenceUnits{15e14} and \fluenceUnits{25e14} are shown in (a), (b), (c) and (d) respectively.}
        \label{fig:IV_irr}
\end{figure}

Following the irradiation of the devices, a second electrical characterization was performed at \SI{-25}{\celsius}, from where two sensors from each carbon dose and fluence are presented in \autoref{fig:IV_irr}. The pad leakage current of the sensors of different carbon doses can be seen as a function of the reverse bias in (a), (b), (c) and (d) for the fluences of \fluenceUnits{4e14}, \fluenceUnits{8e14}, \fluenceUnits{15e14} and \fluenceUnits{25e14} respectively. The displacement of the $V_{BD}$ regimes after irradiation is clearly visible, in the case of the standard samples, starting from about \SI{540}{\volt} for the lowest fluence and about \SI{740}{\volt} for the irradiated at \fluenceUnits{15e14}, and in the case of the carbonated sensors there is not big difference in the $V_{BD}$ between low carbon and high carbon samples, at the lowest fluence their $V_{BD}$ is lower compared to the standard samples but at higher fluences this difference is decreasing. This increase in the $V_{BD}$ indicates the degradation of the gain layer due to irradiation fluence. For the carbonated sensors in plot (c) the noise increased significally before reach the $V_{BD}$ and measures had to be stopped at \SI{600}{\volt} and do not go to higher voltages to avoid damaging them.

\subsection{Capacitance-Voltage characteristic}
\label{sec:cv}

\begin{figure}
     \centering
     \begin{subfigure}[b]{0.49\textwidth}
         \centering
         \includegraphics[width=\textwidth, trim={0.6cm 0.4cm 1.5cm 0.7cm}, clip=true]{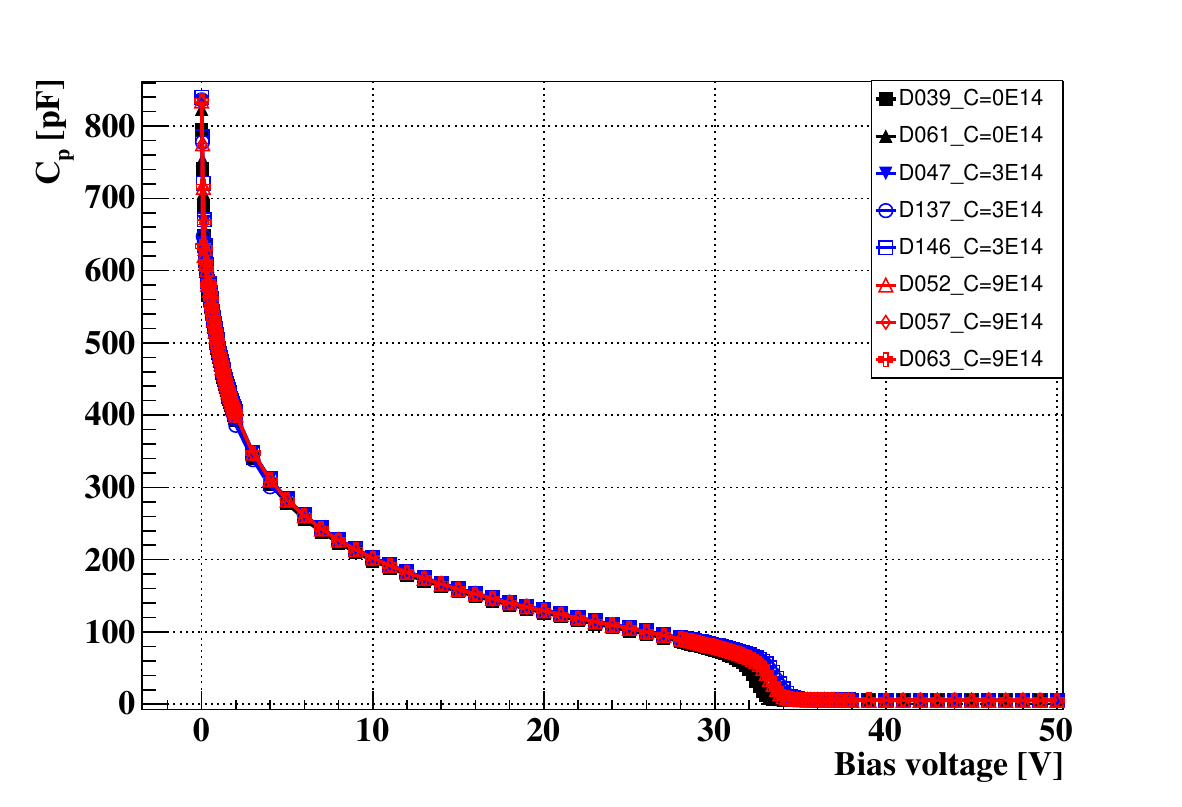}
         \caption{CV curves}
     \end{subfigure}
     \hfill
     \begin{subfigure}[b]{0.49\textwidth}
         \centering
         \includegraphics[width=\textwidth, trim={0.6cm 0.4cm 1.5cm 0.7cm}, clip=true]{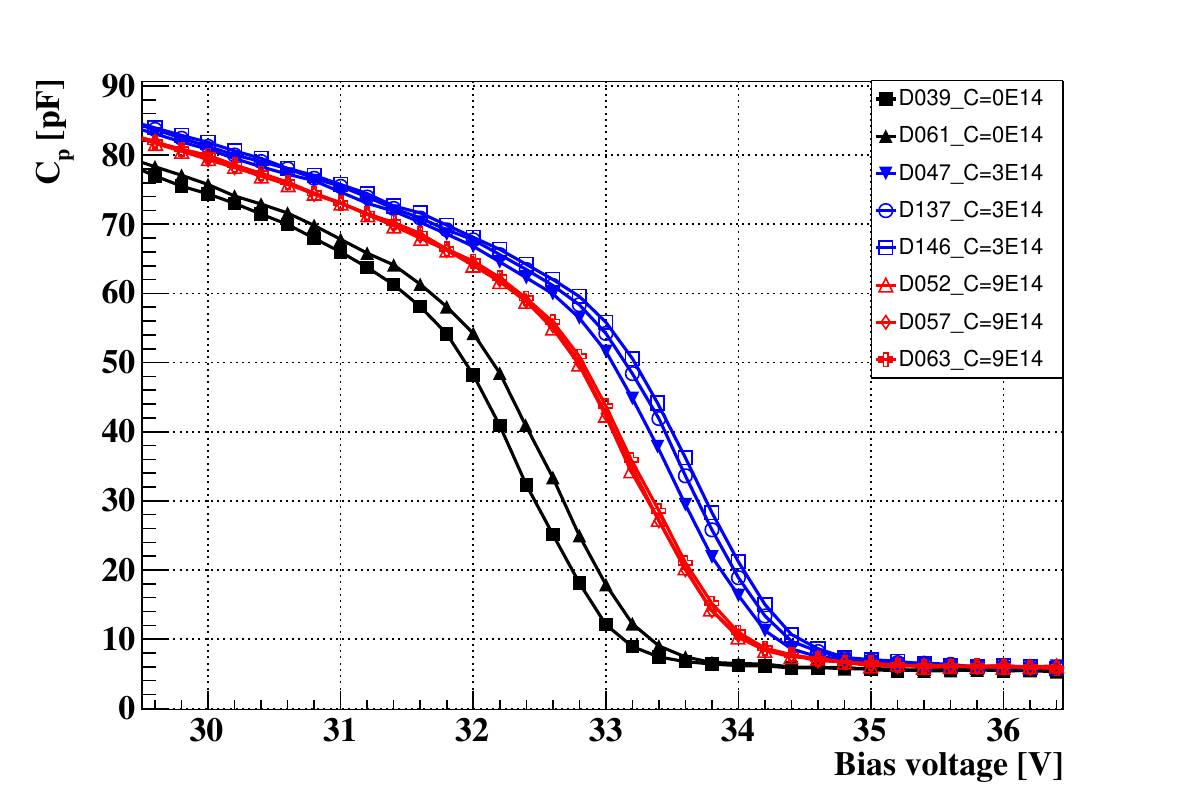}
         \caption{GL region enlarged}
     \end{subfigure}
        \caption{Pad capacitance before irradiation as a function of the reverse bias. The characteristic kinks in the curves due to the gain layer and bulk depletion can be observed.}
        \label{fig:cv_pre}
\end{figure}

Before irradiation, the capacitance of the bare sensors was measured at room temperature using an LCR-meter at a frequency of \SI{100}{\hertz} and with the guard-ring connected. The capacitance curves against the applied reverse bias for both carbonated and standard samples are shown in plot (a) of figure \autoref{fig:cv_pre}. These curves exhibit high uniformity and repeatability. The CV curves start with a quick decrease in capacitance according to the depletion due to the biasing and then continues decreasing smoothly until before the \SI{31}{\volt} for standard samples and  about \SI{32}{\volt} for the carbonated, indicating the depletion voltage of the gain layer \(V_{GL}\). This is followed by another turning point in the curve in which the capacitance decreases quickly, signifying the depletion of the bulk and finally reaching a plato region at a final capacitance \(C_{\text{end}}\) of approximately \SI{4}{\pico\farad} at a bias of \SI{34}{\volt} for standard samples and \SI{35}{\volt} for carbonated samples. This final capacitance is in line with the fact that all sensors share the same dimensions. Plot (b) is an enlarged view of the $V_{GL}$ region from the same measurements, revealing that all samples, whether low carbonated, high carbonated or standard, have similar curve shapes.

\begin{figure}
     \centering
     \begin{subfigure}[b]{0.49\textwidth}
         \centering
         \includegraphics[width=\textwidth, trim={0.6cm 0.4cm 1.5cm 0.7cm}, clip=true]{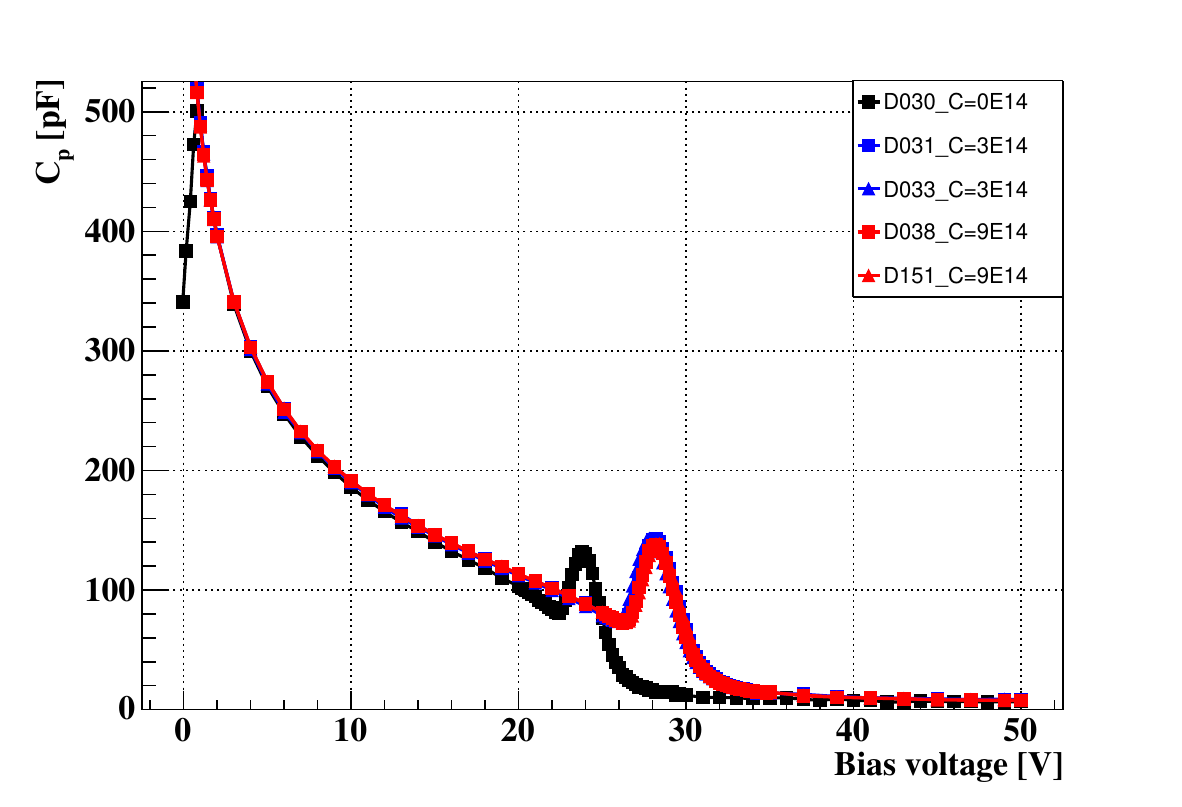}
         \caption{Samples irradiated at \fluenceUnits{4e14}}
     \end{subfigure}
     \hfill
     \begin{subfigure}[b]{0.49\textwidth}
         \centering
         \includegraphics[width=\textwidth, trim={0.6cm 0.4cm 1.5cm 0.7cm}, clip=true]{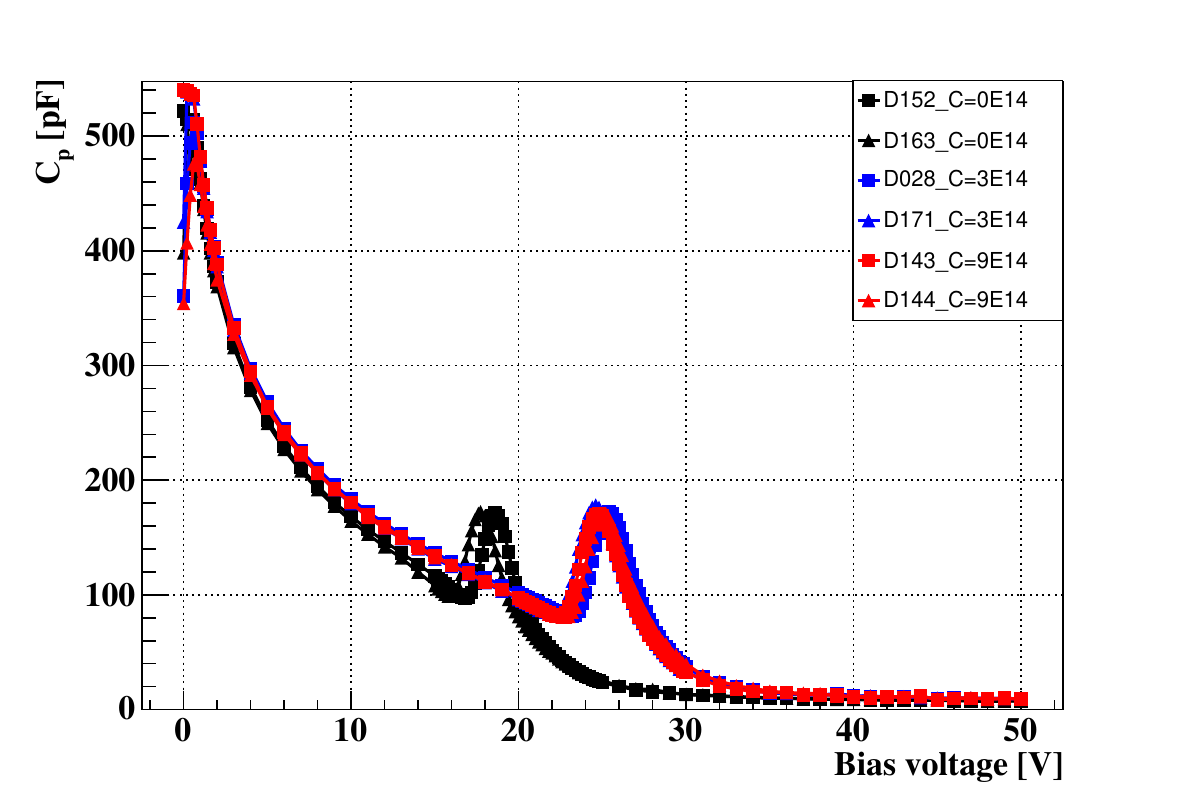}
         \caption{Samples irradiated at \fluenceUnits{8e14}}
     \end{subfigure}
     \begin{subfigure}[b]{0.49\textwidth}
         \centering
         \includegraphics[width=\textwidth, trim={0.6cm 0.4cm 1.5cm 0.7cm}, clip=true]{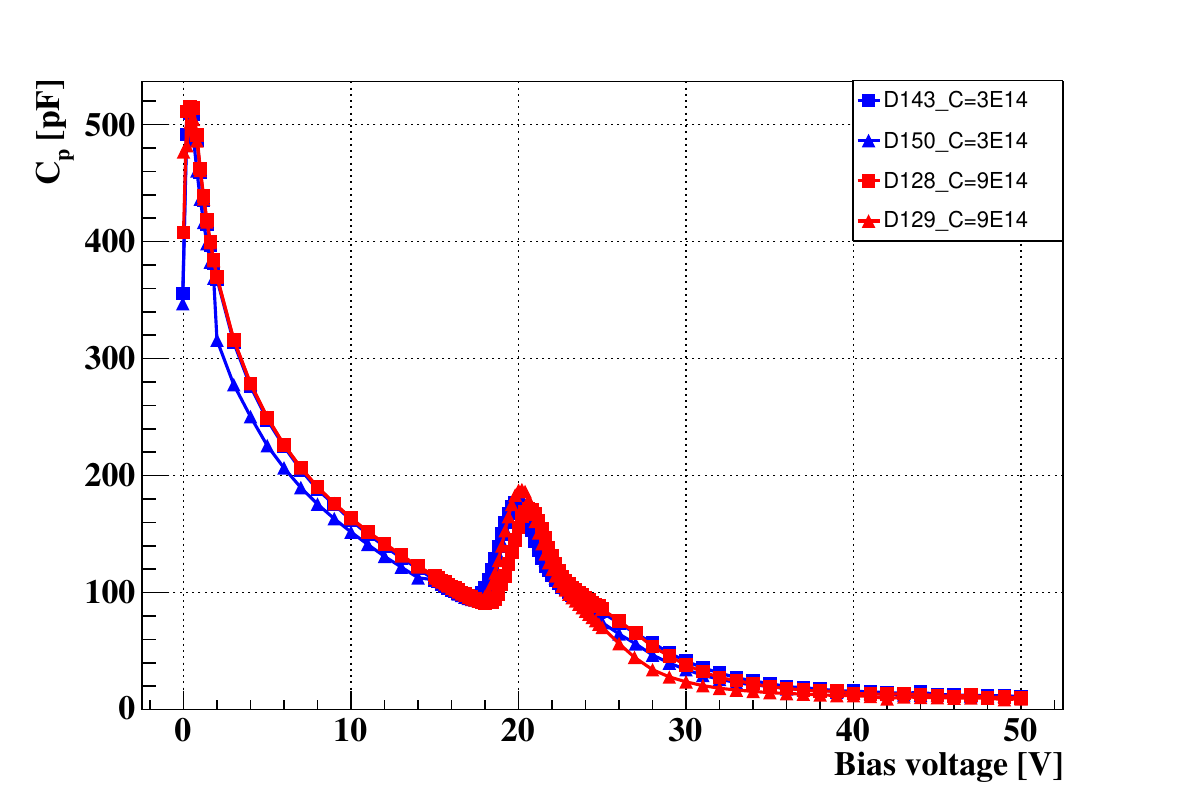}
         \caption{Samples irradiated at \fluenceUnits{15e14}}
     \end{subfigure}
     \hfill
     \begin{subfigure}[b]{0.49\textwidth}
         \centering
         \includegraphics[width=\textwidth, trim={0.6cm 0.4cm 1.5cm 0.7cm}, clip=true]{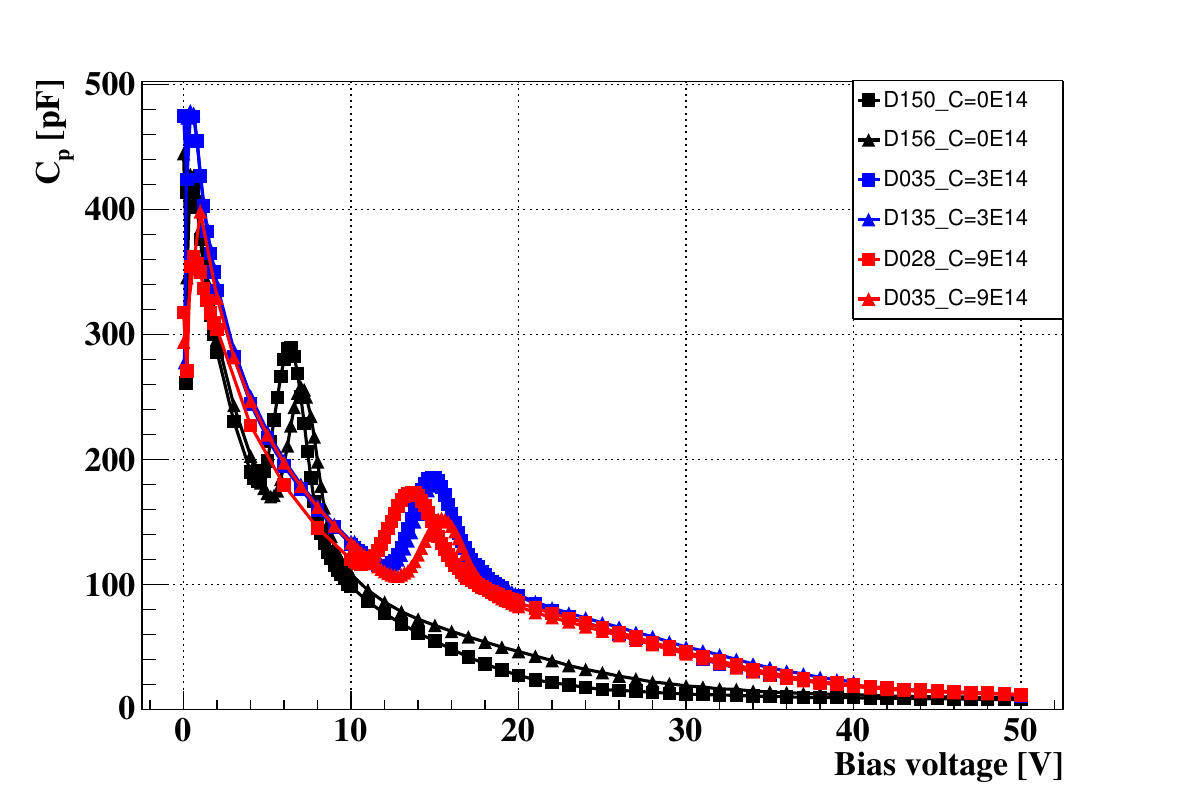}
         \caption{Samples irradiated at \fluenceUnits{25e14}}
     \end{subfigure}

        \caption{Pad capacitance after irradiation as a function of the reverse bias. Standard and carbonated devices are shown in (a), (b), (c) and (d), according to the fluence points: \fluenceUnits{4e14}, \fluenceUnits{8e14}, \fluenceUnits{15e14} and \fluenceUnits{25e14}. Displacement of the $V_{GL}$ (start of the peak in the curve) as result of the irradiation at the four fluence points is observed.}
        \label{fig:cv_after}
\end{figure}

After the irradiation of the samples, according to previous studies~\cite{Campbell}, in order to observe the $V_{GL}$ region from the CV curves in a better way, the measurements were performed in the same configuration as before irradiation with a low frequency of \SI{100}{\hertz} in the LCR-meter and at a temperature of \SI{10}{\celsius}. The pad capacitance of the samples after irradiation can be observe as CV curves in~\autoref{fig:cv_after}. In these CV curves of the irradiated samples, we can observe that after the initial decrease of the capacitance, it starts to increase again until reach a local maximum, which is an effect of the presence of the gain layer according to other studies~\cite{Moritz} and seen before in other irradiated samples studies. 

Starting with the point that all irradiated samples measurements have a similar behaviour corresponding to the gain layer region, we can observe that there are two main differences across all samples: The first one comes from the irradiation fluence that leads to a shift in the $V_{GL}$ that is lower at higher fluences, being the $V_{GL}$ higher at a lower fluence. This can be understood with the fact that the bias voltage required to deplete the gain layer is less at higher degradation in the gain layer due to irradiation. The second difference to observe is the effect of the carbon, since in the plots (except plot (c) in which the standard samples could not be measured), the carbonated samples has higher $V_{GL}$ compared with the standard samples but between the carbonated samples the difference is less evident, taking for example plot (a) in which $V_{GL}$ of carbonated devices is around \SI{26.5}{\volt} while for the standard sensor it is around \SI{22}{\volt}. We have determined the $V_{GL}$ for these measurements as the voltage of the lower capacitance before the rise in the curve, since this value agrees with the $V_{GL}$ from IV curves.

\subsection{Acceptor Removal Coefficient Determination}
\label{sec:acceptor}

Research has demonstrated that Low Gain Avalanche Detectors (LGAD) sensors undergo a decrease in gain following irradiation with charged hadrons or neutrons~\cite{Galloway2019}. This decrease can be ascribed to the initial acceptor removal mechanism, which involves the progressive deactivation of acceptors that constitute the Gain Layer (GL)~\cite{Kramberger}, particularly Boron (B) in this study. 

As the irradiation process deactivates the Boron implanted in the GL of the devices, the bias voltage needed to completely deplete this gain layer, diminishes compared to its pre-irradiation state. This decrease in $V_{GL}$ serves as an indicator of the residual active Boron in the GL. Given the assumption of uniform Boron removal across the multiplication layer at a steady rate, $V_{GL}$ can be expressed as being proportional to the Boron concentration using the equation below:

\begin{equation}
\label{eq:acceptor}
V_{GL}(\Phi)\approx V_{GL}(\Phi=0)\times \exp^{-c\Phi}
\end{equation}

In this equation, $c$ denotes the acceptor removal coefficient, and $V_{GL}$ signifies the gain layer depletion voltage corresponding to the specified fluence $\Phi$. The coefficient $c$ serves as a measure of the degradation experienced by the multiplication layer, implying that a lower $c$ value indicates a more radiation-hardened sensor.

As seen in \autoref{sec:Electric}, a subsequent electrical characterization after irradiation was performed to examine the degradation of the gain layer, beginning with the calculation of the $V_{GL}$ and then the determination of the Acceptor Removal Coefficient. A key element of this study involves investigating the impact of different doses of carbon enrichment in the GL in comparison to standard Boron implantation, and how it affects the acceptor removal coefficient for the different samples. The $V_{GL}$ values extracted from the electrical characterization are presented in ~\autoref{tab:vgls}. These values allow for the calculation of the degradation of the GL by fitting the dependence of $V_{GL}$ with fluence, according to~\autoref{eq:acceptor}.

\begin{table}[htbp]
\centering
\caption{\label{tab:vgls} Summary of the $V_{GL}$ values for both type of sensors, extracted from the electrical characterization (CV) before and after irradiation. The errors are the standard error of the mean (SEM) from the samples measured.}
\smallskip
\begin{tabular}{lccc}
\hline
\multicolumn{4}{c}{$V_{GL}$ from CV (V)} \\

Fluence ($\fluenceUnits{ }$) & Standard & Low C & High C \\
\hline\hline
 0 & $31.6\pm.2$ & $32.7\pm.2$ & $32.5\pm.2$  \\

$0.4\times 10^{15}$ & $22.3\pm.2$ & $25.9\pm.2$  & $27.1\pm.2$   \\

$0.8\times 10^{15}$ & $16.3\pm.2$ & $22.6\pm.2$  & $24.0\pm.2$    \\

$1.5\times 10^{15}$ & - & $17.2\pm.2$ & $18.9\pm.2$    \\

$2.5\times 10^{15}$ & $4.8\pm.2$ & $17.2\pm.2$  & $12.1\pm.2$  \\
\hline
\end{tabular}
\end{table}

\begin{figure}
     \centering
     \begin{subfigure}[b]{0.7\textwidth}
         \centering
         \includegraphics[width=\textwidth, trim={0.6cm 0cm 1.5cm 0.7cm}, clip=true]{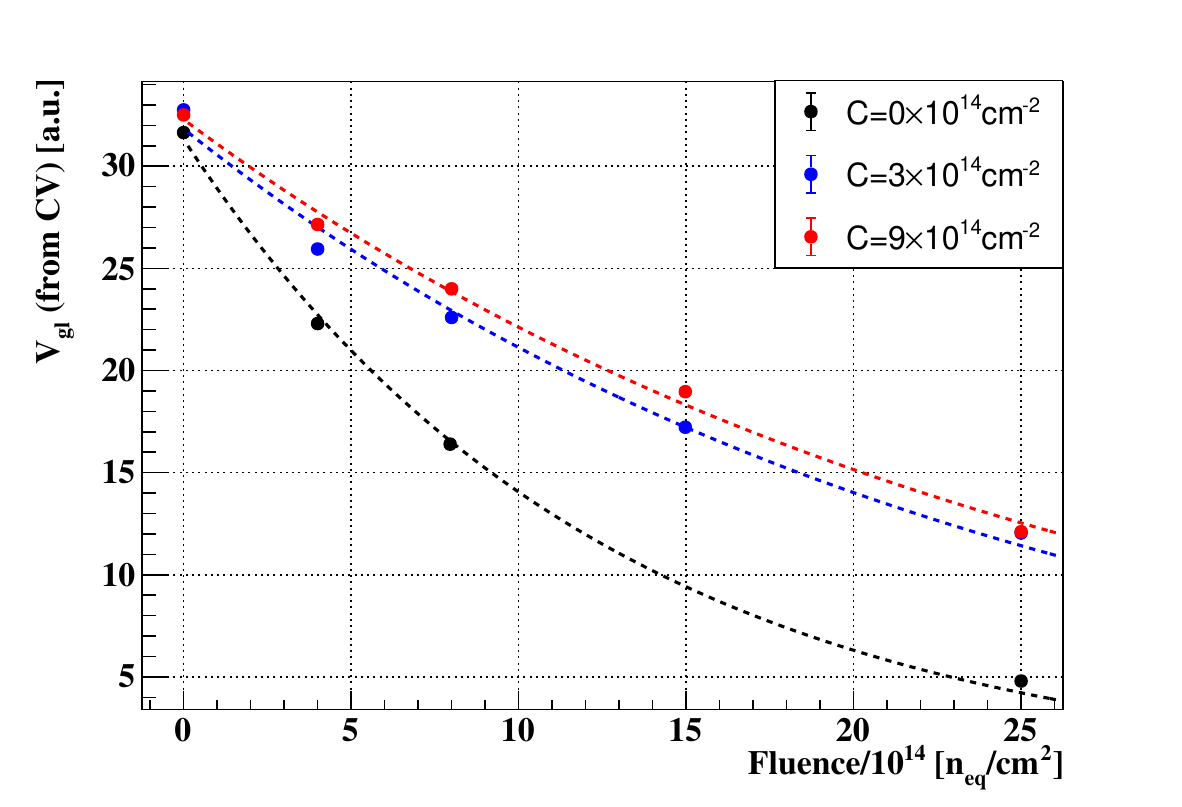}
     \end{subfigure}
     \hfill
         \caption{Gain-layer depletion voltage from CV as a function of the fluence and the respective fit (doted lines) from where the acceptor removal coefficients are calculated and being $c[\SI{e-16}{\centi\meter\squared}]=7.9$ for the standard samples, $c[\SI{e-16}{\centi\meter\squared}]=3.5$ for the samples with $C=\SI{3e14}{\centi\meter\squared}$ and $c[\SI{e-16}{\centi\meter\squared}]=4.4$ for the samples with $C=\SI{9e14}{\centi\meter\squared}$.}
        \label{fig:Acceptor}
\end{figure}

\autoref{fig:Acceptor} shows the resulting curves for the $V_{GL}$ versus fluence for all the measured samples carbonated and standard and the fits for these curves in order to calculate the acceptor removal coefficients. The resulting coefficients are $c[\SI{e-16}{\centi\meter\squared}] = 7.9$ for the standard samples, $c[\SI{e-16}{\centi\meter\squared}]=3.5$ for the samples with $C=\SI{3e14}{\centi\meter\squared}$ and $c[\SI{e-16}{\centi\meter\squared}]=4.4$ for the samples with $C=\SI{9e14}{\centi\meter\squared}$. These acceptor removal coefficients indicates that the devices with carbonated GL suffer less degradation and then less Boron deactivation allowing a better radiation tolerance on these samples.

\section{Beta Source Characterization}
\label{sec:RS}
The IFCA’s radioactive source setup consists in a Faraday cage that contains a stack of three sensors, being the bottom sensor always a non-irradiated sample to serve as a reference. Each sensor is affixed to a basic passive PCB that facilitates electrical connections of the devices. This cage is situated within a climate chamber to control the characterization temperature. The beta source is an encapsulated Sr$^{90}$ radioactive source, with an activity of \SI{3.7}{\mega\becquerel}, this is positioned atop the stack, ensuring there’s no direct contact with the samples. The sensors are aligned using mechanical templates to fix them inside the stack structure. An external low-noise current amplifier, with a standard gain of \SI{40}{\decibel}~\cite{CIVIDEC}, is employed to measure the induced current in every sample. An oscilloscope, with a sampling rate of \SI{5}{\giga S\per\second}, is used for readout, which is triggered by a triple coincidence from the stack and recorded as an event, and thousands of events are taken per every bias voltage applied to the samples. The samples measured in the radioactive source setup are outlined in \autoref{tab:rs-sensors}.

\subsection{Collected Charge}
\label{sec:Charge}

\begin{figure}
     \centering
     \begin{subfigure}[b]{0.49\textwidth}
         \centering
         \includegraphics[width=\textwidth, trim={0.6cm 0.4cm 1.5cm 0.7cm}, clip=true]{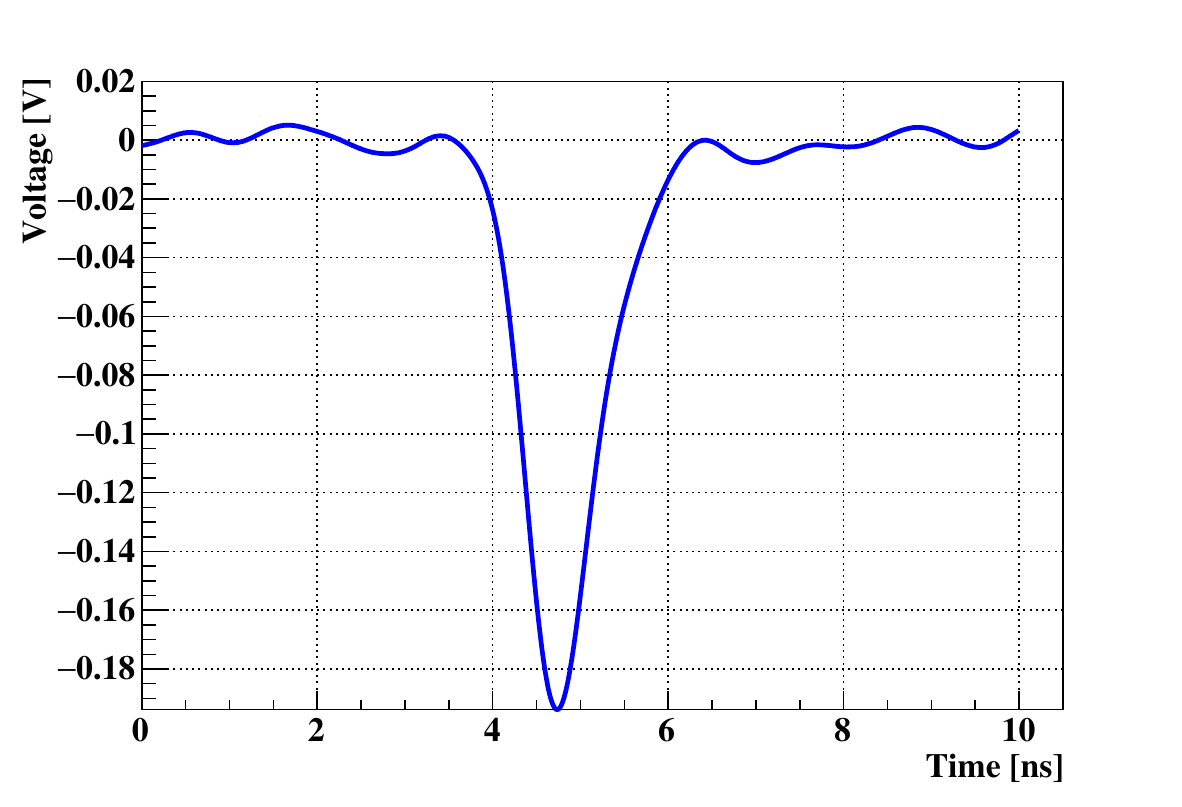}
         \caption{Typical waveform.}
     \end{subfigure}
     \hfill
     \begin{subfigure}[b]{0.49\textwidth}
         \centering
         \includegraphics[width=\textwidth, trim={0.6cm 0.4cm 1.5cm 0.7cm}, clip=true]{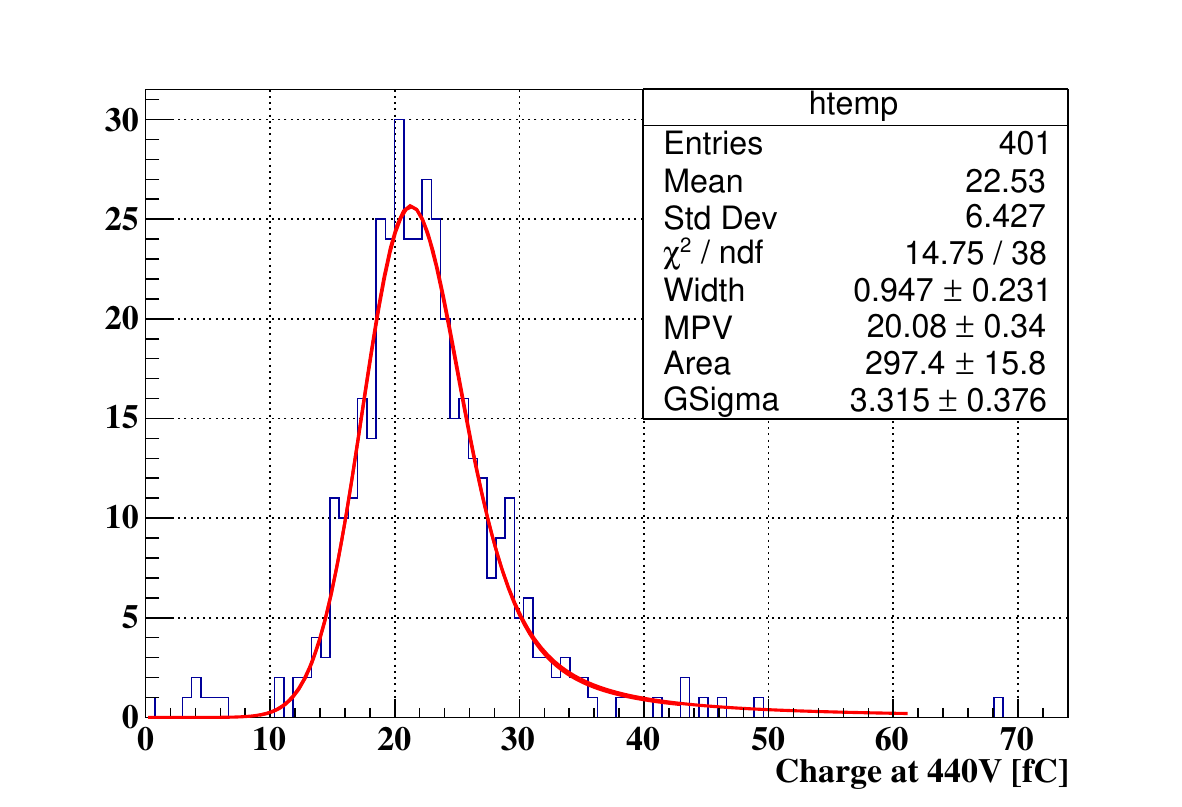}
         \caption{Collected charge.}
     \end{subfigure}
        \caption{Plot (a) show the waveform from an LGAD: voltage response versus the time of the pulse from a non-irradiated LGAD biased at $\SI{260}{\volt}$ . Plot (b) is the distribution of the collected charge computed from the integration of the waveforms of an carbonated LGAD irradiated to $\fluenceUnits{15e14}$ and biased at $\SI{440}{\volt}$. The Most Probable Value (MVP) can be extracted from the convoluted Gauss-Landau fit.}
        \label{fig:chargeColl}
\end{figure}

The charge collected is computed as the integral of the voltage pulse as shown in \autoref{fig:chargeColl} (a). The total charge distribution for a single detector, depicted in \autoref{fig:chargeColl} (b), then is fitted by convoluting a Landau with a Gaussian from where we can extract the Most Probable Value (MPV) of this distribution as the total collected charge.

\begin{figure}
     \centering
     \begin{subfigure}[b]{0.49\textwidth}
         \centering
         \includegraphics[width=\textwidth, trim={0.6cm 0.4cm 1.5cm 0.7cm}, clip=true]{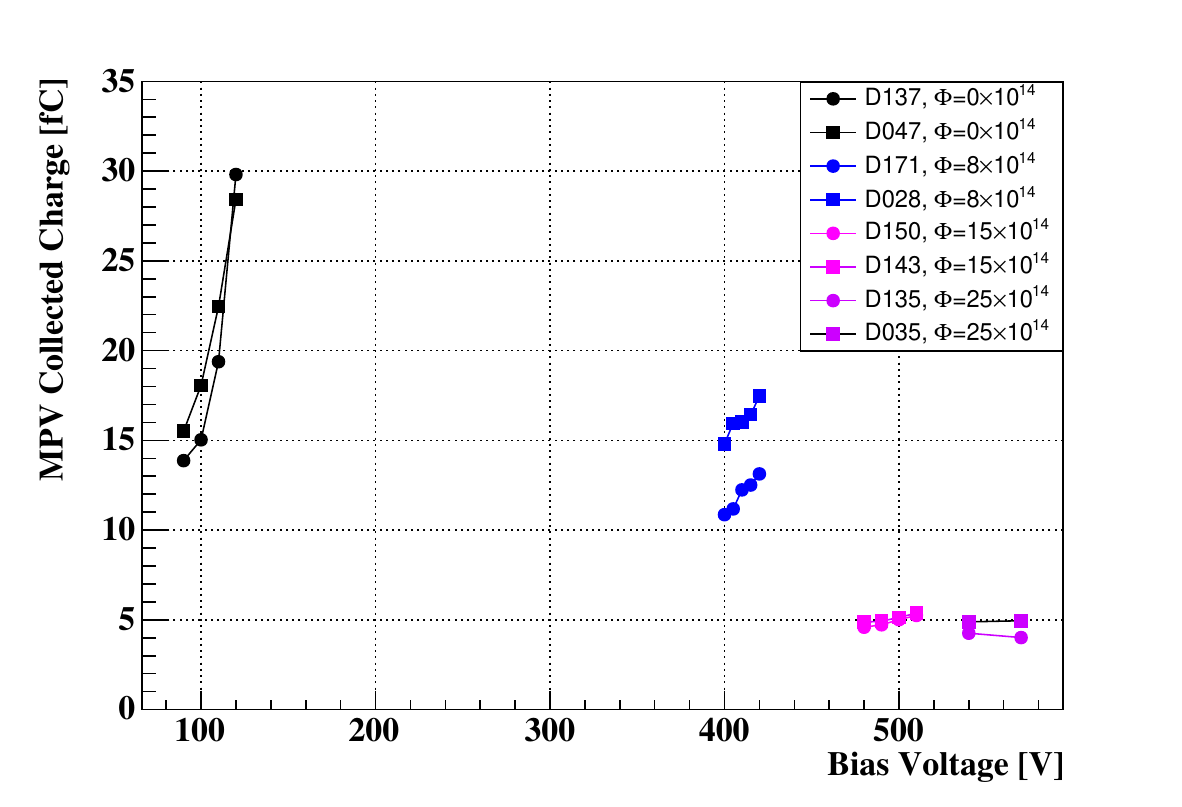}
         \caption{Low Carbonated}
     \end{subfigure}
     \hfill
     \begin{subfigure}[b]{0.49\textwidth}
         \centering
         \includegraphics[width=\textwidth, trim={0.6cm 0.4cm 1.5cm 0.7cm}, clip=true]{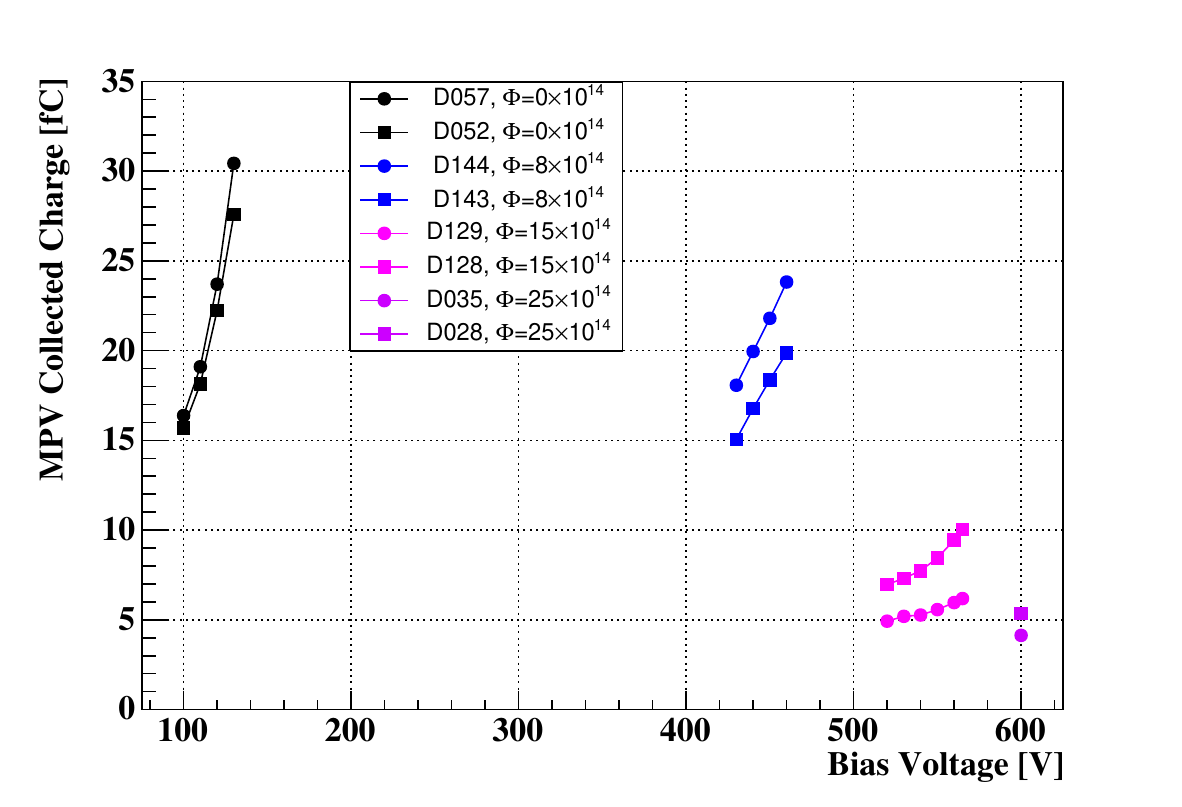}
         \caption{High Carbonated}
     \end{subfigure}
        \caption{Plots of the Collected Charge as a function of the reverse bias voltage for low carbonated samples (a) and high carbonated samples (b). There is a difference in the voltage required to achieve the same level of collected charge between the two types of LGADs at a same fluence value. All these measurements were performed at -25C}
        \label{fig:Charge}
\end{figure}

The collected charge as a function of the bias voltage applied to the samples from different fluences are shown in \autoref{fig:Charge} (a) containing the low carbonated samples and (b) the high carbonated. As expected, the collected charge between the samples of the same fluence and carbon dose are close to each other, being evident the effect of the radiation since the bias voltage required to collected a certain amount of charge is higher for the more irradiated samples. For comparison, low carbonated samples irradiated to \fluenceUnits{1.5e15} have a collected charge of \SI{5}{\femto\coulomb} at \SI{500}{\volt} and the samples of \fluenceUnits{2.5e15} require more than \SI{540}{\volt} instead. Another thing to mention is that there is not so much difference between low and high carbonated samples in terms of the collected charge if applying the same bias voltage, but some low carbonated samples could not be operated to higher voltage like the high carbonated, due to the presence of noise particularly at the higher fluence.

\subsection{Time Resolution}
\label{sec:time}

The time resolution of these devices, which can be determined as the standard deviation of the distribution of the difference in the sensor’s time of arrival (ToA) relative to a time reference sensor, is a crucial parameter. Typically, a well-known detector serves as this time reference. When three detectors of unknown characteristics are measured concurrently, their individual time resolutions can be derived from the three relative differences~\cite{McKarris} that we can denote as (1-2, 1-3, 2-3) according to the position of the sensors inside the stack described in \autoref{sec:RS}.

The ToA is computed as the instant a pulse surpasses a certain threshold. Due to the time walk effect, where pulses of varying amplitudes that arrive simultaneously cross a threshold at different times, a Constant Fraction Discrimination (CFD) algorithm is employed to correct the pulses. 

By recording these ToAs from the three channels (one per sample), we can determine the time difference among the three sensors, then the fitted widths: $\sigma_{1,2}$, $\sigma_{1,3}$, and $\sigma_{2,3}$ of the distributions of this differences are used to calculate the time resolutions ($\sigma_{1}$, $\sigma_{2}$, and $\sigma_{3}$) corresponding to each sample by using \autoref{eq:sigmas}:

\begin{equation}
\label{eq:sigmas}
\begin{split}
\sigma_1= \left( \frac{1}{2} (\sigma_{2,1}^{2}+\sigma_{1,3}^{2}-\sigma_{3,2}^{2})\right)^{\frac{1}{2}} \,, \\
\sigma_2= \left(\frac{1}{2} (\sigma_{2,1}^{2}-\sigma_{1,3}^{2}+\sigma_{3,2}^{2})\right)^{\frac{1}{2}} \,, \\
\sigma_3= \left(\frac{1}{2} (-\sigma_{2,1}^{2}+\sigma_{1,3}^{2}+\sigma_{3,2}^{2})\right)^{\frac{1}{2}} \,, 
\end{split}
\end{equation}

and its errors ($\delta_{1}$, $\delta_{2}$ and $\delta_{3}$) from \autoref{eq:deltas}:

\begin{equation}
\label{eq:deltas}
\begin{split}
\delta_1= \frac{ \left( (\sigma_{2,1} \delta_{2,1})^2 + (\sigma_{1,3} \delta_{1,3})^2 + (\sigma_{3,2} \delta_{3,2})^2\right)^{\frac{1}{2}} }{2\sigma_1}\,,\\
\delta_2= \frac{ \left( (\sigma_{2,1} \delta_{2,1})^2 + (\sigma_{1,3} \delta_{1,3})^2 + (\sigma_{3,2} \delta_{3,2})^2\right)^{\frac{1}{2}} }{2\sigma_2}\,,\\
\delta_3= \frac{ \left( (\sigma_{2,1} \delta_{2,1})^2 + (\sigma_{1,3} \delta_{1,3})^2 + (\sigma_{3,2} \delta_{3,2})^2\right)^{\frac{1}{2}} }{2\sigma_3}\,,
\end{split}
\end{equation}

where $\delta_{i,j}$ is the error in the value $\sigma_{i,j}$.

\begin{figure}
     \centering
     \begin{subfigure}[b]{0.49\textwidth}
         \centering
         \includegraphics[width=\textwidth, trim={0.6cm 0.4cm 1.5cm 0.7cm}, clip=true]{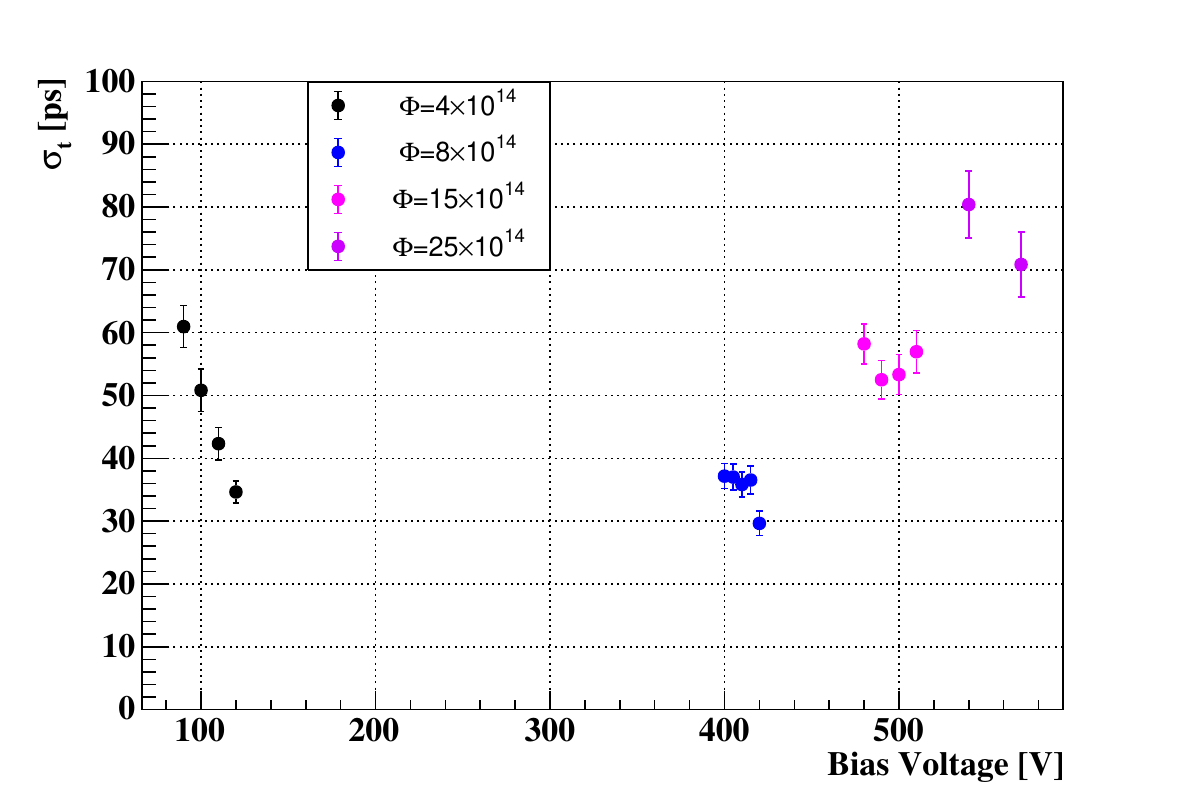}
         \caption{Low Carbonated}
     \end{subfigure}
     \hfill
     \begin{subfigure}[b]{0.49\textwidth}
         \centering
         \includegraphics[width=\textwidth, trim={0.6cm 0.4cm 1.5cm 0.7cm}, clip=true]{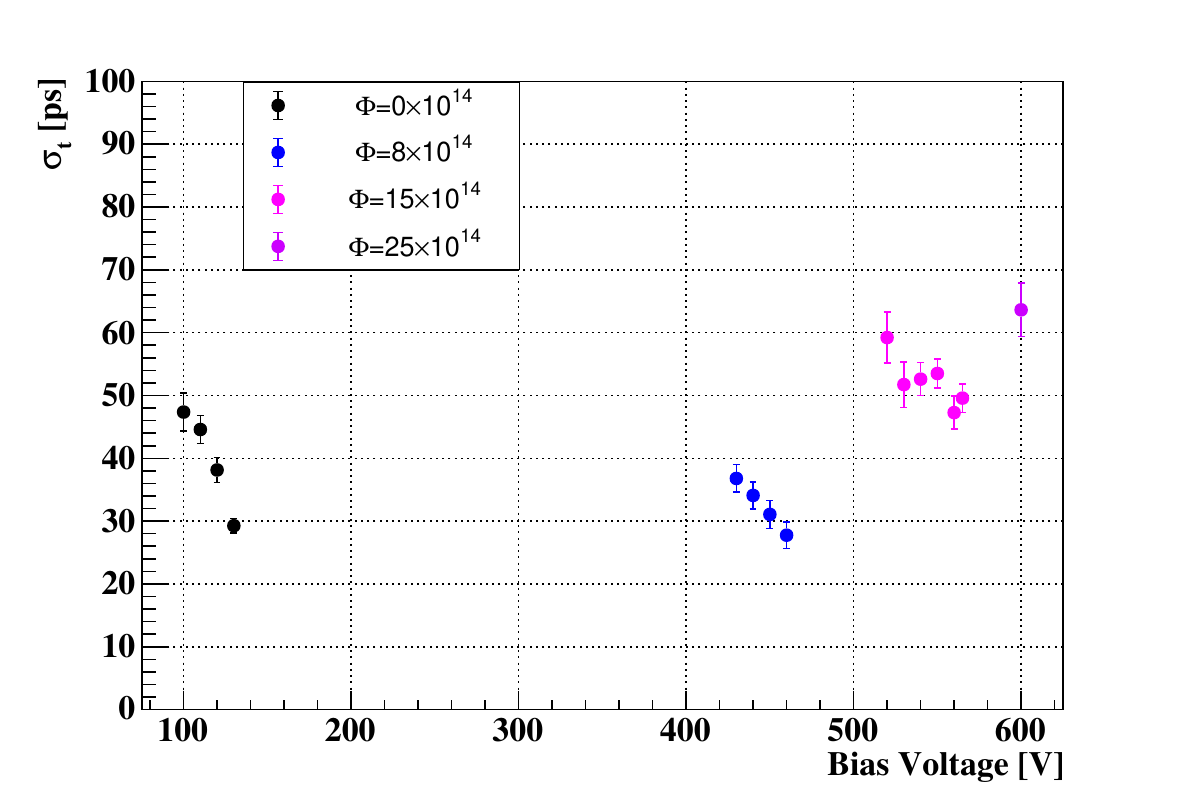}
         \caption{High Carbonated}
     \end{subfigure}
        \caption{Time resolution of the both type of sensors: Low carbonated (a) and high carbonated (b), calculated using \autoref{eq:sigmas} and errors with \autoref{eq:deltas}. All these measurements were performed at $\SI{-25}{\celsius}$}.
        \label{fig:Time}
\end{figure}

The procedure for determining the time resolution was replicated across all the samples in this study, with the non-irradiated sensor referenced in \autoref{sec:RS} serving as the time reference. The resultant time resolutions $\sigma_t$, plotted as a function of the bias voltage, are depicted in \autoref{fig:Time} for both low (a) and high (b) carbonated sensors. It is once again noticeable that as the fluence escalates, the voltage required to obtain an equivalent time resolution also rises, and the time resolution enhances as the bias voltage increases. The both type of sensors can reach a time resolution below \SI{50}{\pico\second} at a fluence of \fluenceUnits{1.5e15}, but as mentioned in the previous section, the low carbonated sensors could not be biased due to the noise while the high carbonated ones could and sensors from the higher fluence point were not able to reach the same resolution values.

\section{Spurious Pulses Study}
\label{sec:noise}

Another important study that leads to understand the correct functionality of the sensors at the operational bias voltage is a noise study, that was conducted on the carbonated samples. This study considered the presence and frequency of micro-discharges that may manifest in silicon detectors as thermally generated spurious pulses. The same characterization setup as described in \autoref{sec:RS} was used, but without the radioactive source shooting with the aim of measure only spurious pulses. These spurious pulses appeared in all the samples near the breakdown voltage, but we decided not to operate them at higher voltages to prevent Single Event Burnout (SEB)~\cite{SEB}.

\begin{figure}
     \centering
     \begin{subfigure}[b]{0.49\textwidth}
         \centering
         \includegraphics[width=\textwidth, trim={0.6cm 0.4cm 1.5cm 0.7cm}, clip=true]{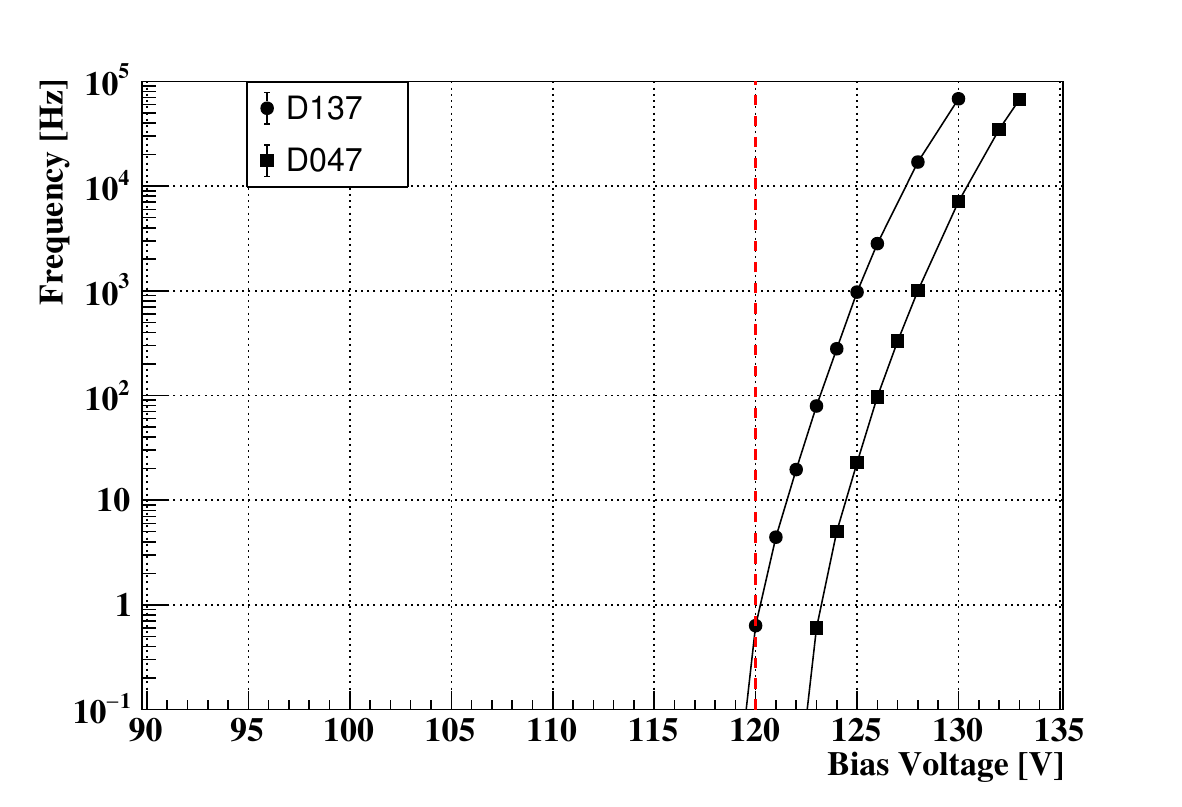}
         \caption{Non-irradiated}
     \end{subfigure}
     \hfill
     \begin{subfigure}[b]{0.49\textwidth}
         \centering
         \includegraphics[width=\textwidth, trim={0.6cm 0.4cm 1.5cm 0.7cm}, clip=true]{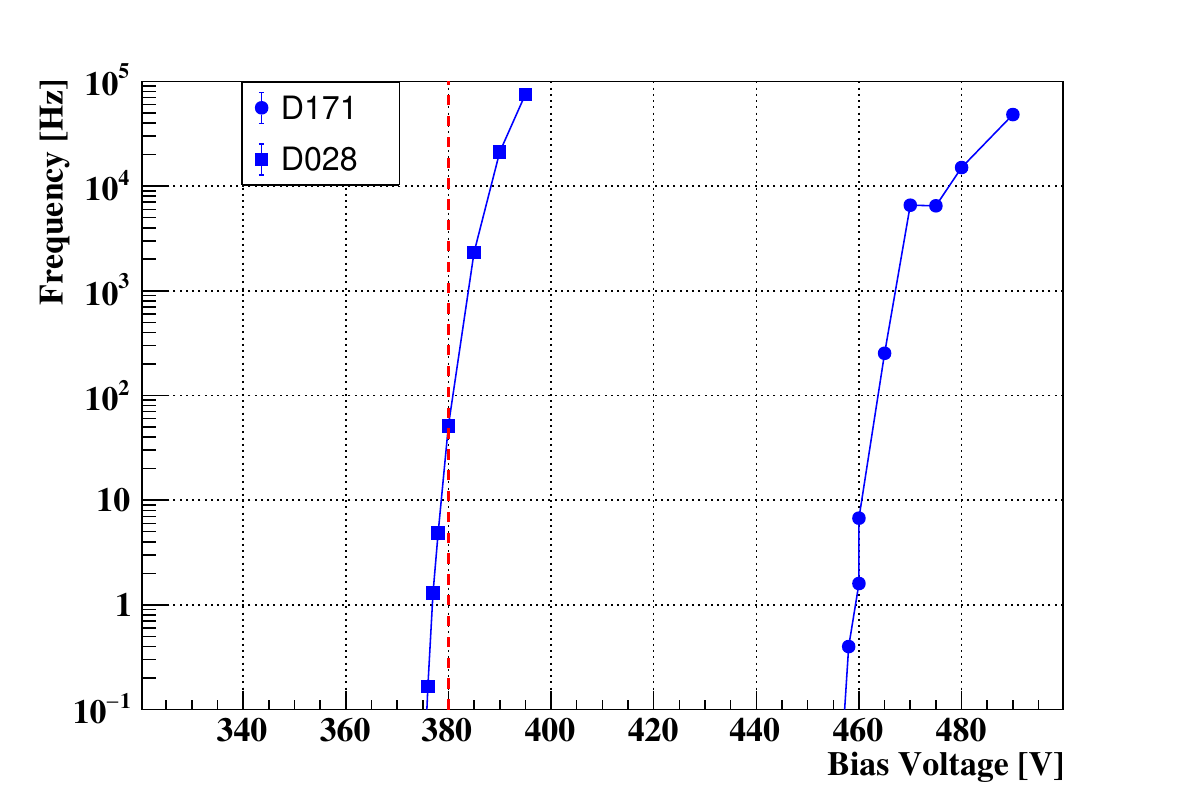}
         \caption{$\fluenceUnits{0.8e15}$}
     \end{subfigure}
     \begin{subfigure}[b]{0.49\textwidth}
         \centering
         \includegraphics[width=\textwidth, trim={0.6cm 0.4cm 1.5cm 0.7cm}, clip=true]{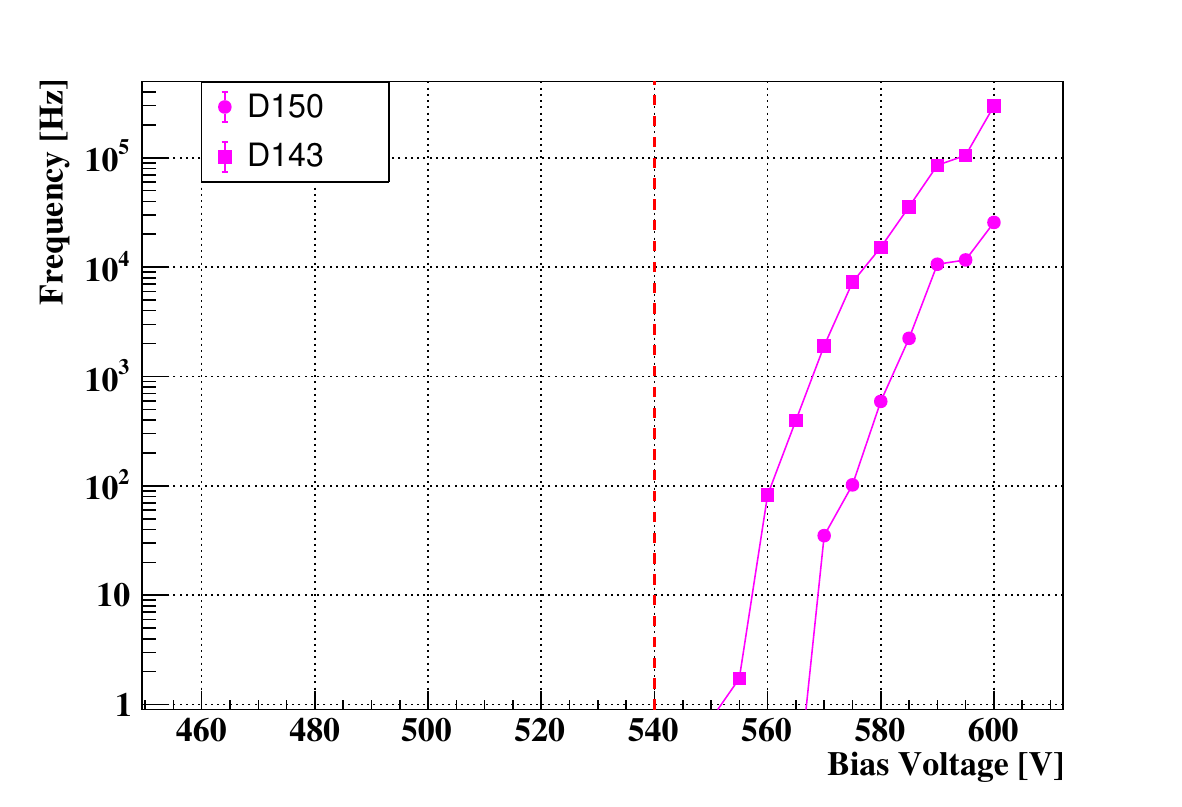}
         \caption{$\fluenceUnits{1.5e15}$}
     \end{subfigure}
     \hfill
     \begin{subfigure}[b]{0.49\textwidth}
         \centering
         \includegraphics[width=\textwidth, trim={0.6cm 0.4cm 1.5cm 0.7cm}, clip=true]{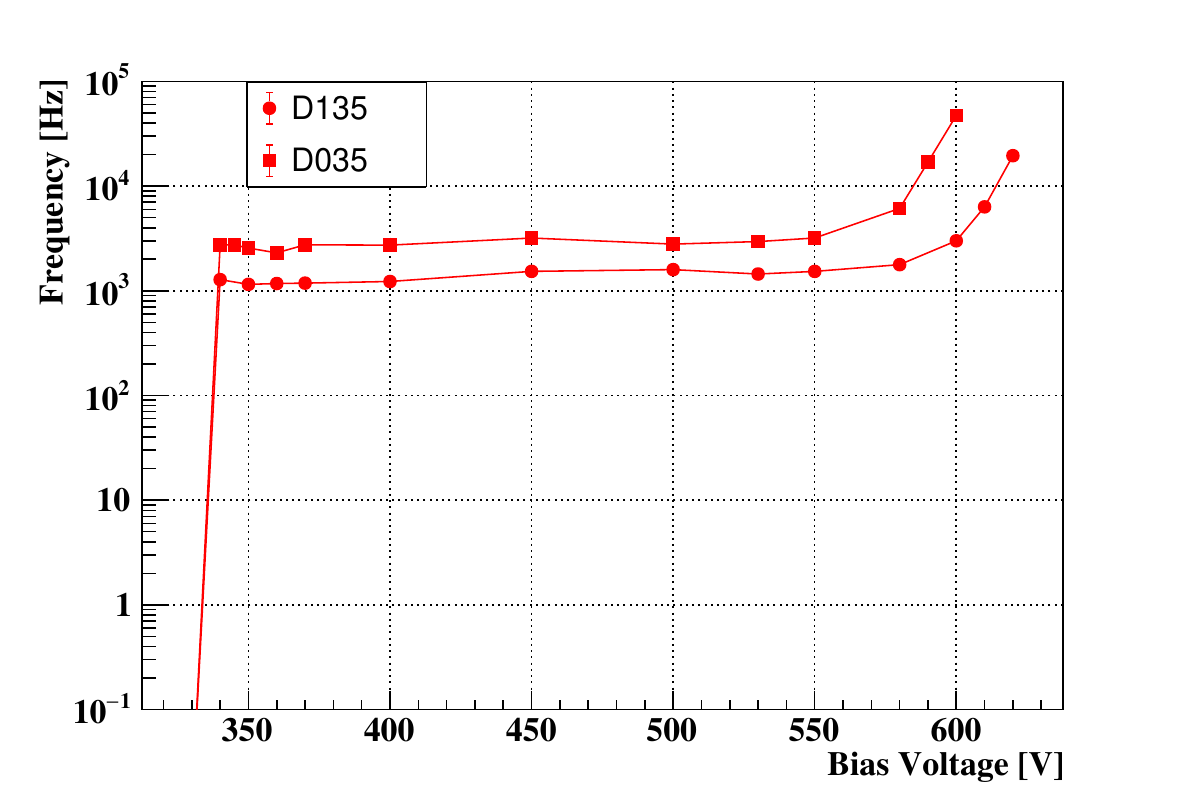}
         \caption{$\fluenceUnits{2.5e15}$}
     \end{subfigure}
        \caption{Spurious pulse rate versus the bias voltage of the Low Carbonated samples when fresh (a), and at $\fluenceUnits{0.8e15}$ (b), $\fluenceUnits{1.5e15}$ (c) and $\fluenceUnits{2.5e15}$ (d) irradiation fluences. Doted lines indicates the working voltage. Measurements taken in the Radioactive Source setup with NIM electronics with a threshold of $\SI{-25}{\milli\volt}$.}
        \label{fig:SpuriousPulseRatesW4}
\end{figure}

\begin{figure}
     \centering
     \begin{subfigure}[b]{0.49\textwidth}
         \centering
         \includegraphics[width=\textwidth, trim={0.6cm 0.4cm 1.5cm 0.7cm}, clip=true]{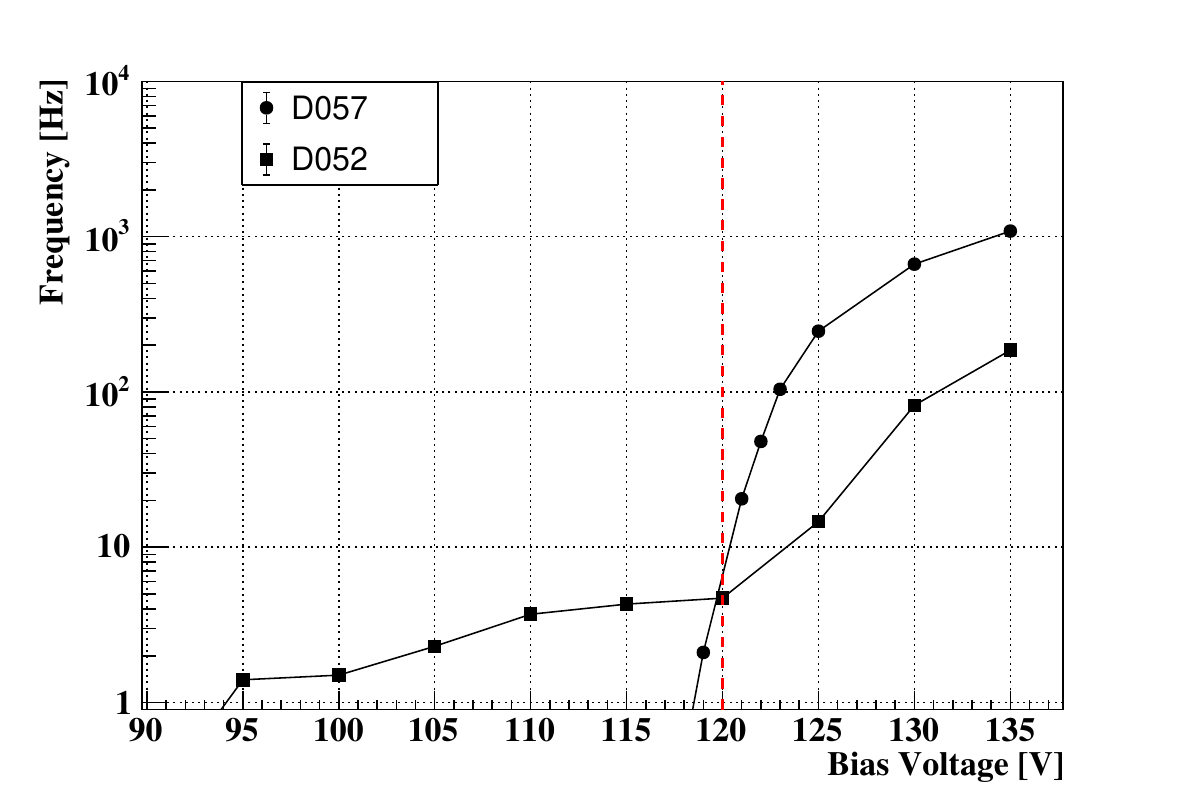}
         \caption{Non-irradiated}
     \end{subfigure}
     \hfill
     \begin{subfigure}[b]{0.49\textwidth}
         \centering
         \includegraphics[width=\textwidth, trim={0.6cm 0.4cm 1.5cm 0.7cm}, clip=true]{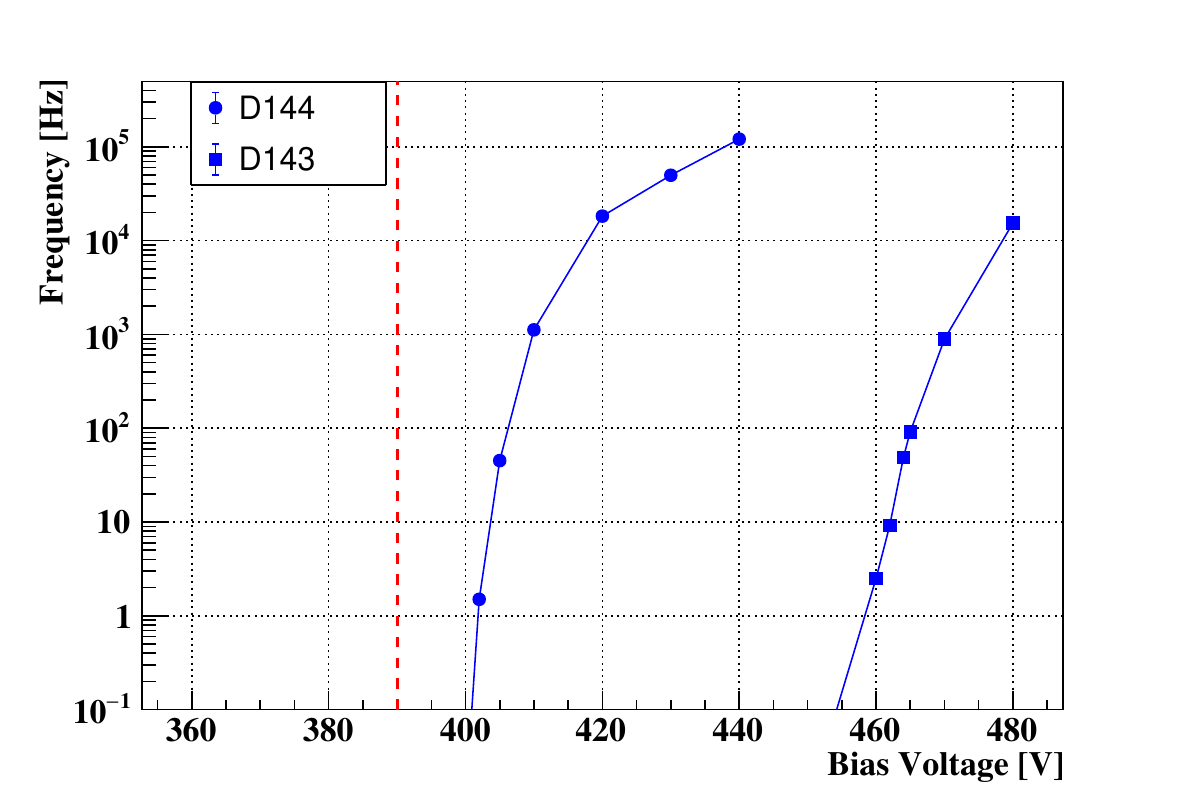}
         \caption{$\fluenceUnits{0.8e15}$}
     \end{subfigure}
     \begin{subfigure}[b]{0.49\textwidth}
         \centering
         \includegraphics[width=\textwidth, trim={0.6cm 0.4cm 1.5cm 0.7cm}, clip=true]{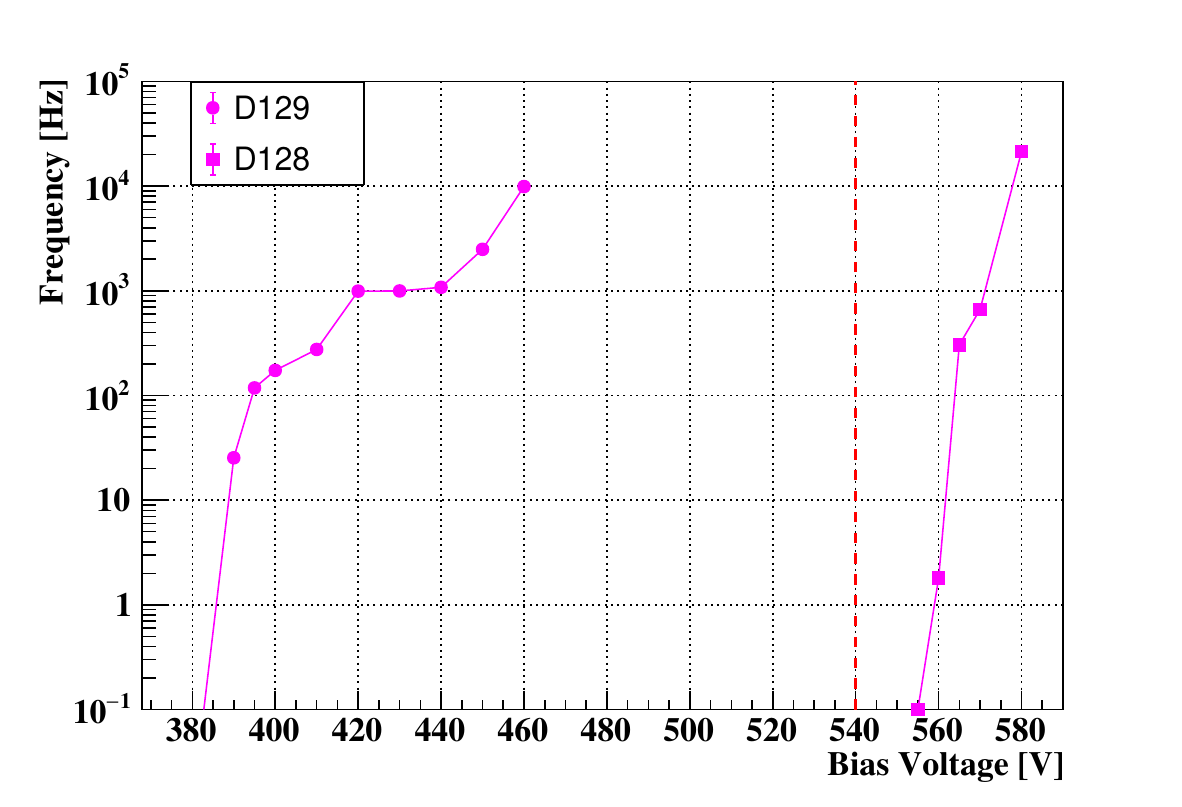}
         \caption{$\fluenceUnits{1.5e15}$}
     \end{subfigure}
     \hfill
     \begin{subfigure}[b]{0.49\textwidth}
         \centering
         \includegraphics[width=\textwidth, trim={0.6cm 0.4cm 1.5cm 0.7cm}, clip=true]{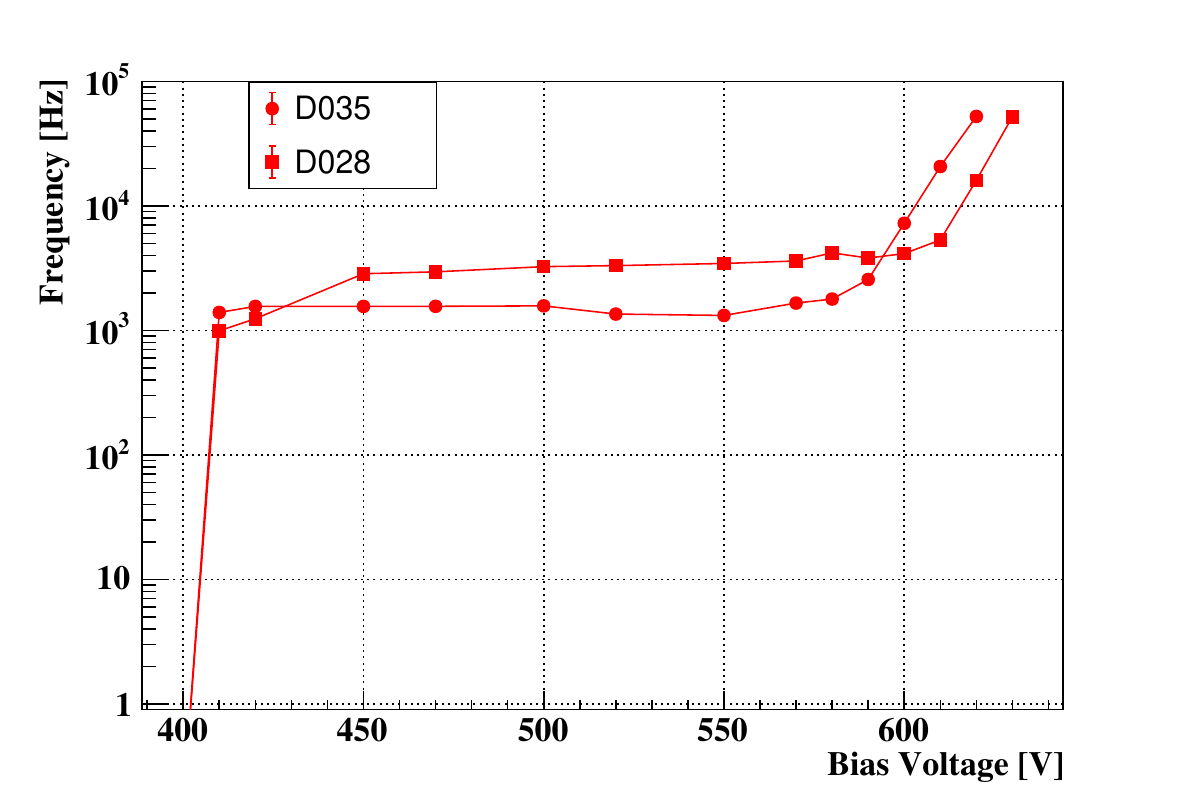}
         \caption{$\fluenceUnits{2.5e15}$}
     \end{subfigure}
        \caption{Spurious pulse rate versus the bias voltage of the High Carbonated samples when fresh (a), and at $\fluenceUnits{0.8e15}$ (b), $\fluenceUnits{1.5e15}$ (c) and $\fluenceUnits{2.5e15}$ (d) irradiation fluences. Doted lines indicates the working voltage. Measurements taken in the Radioactive Source setup with NIM electronics with a threshold of $\SI{-25}{\milli\volt}$.}
        \label{fig:SpuriousPulseRatesW6}
\end{figure}

For this measurements, we decided to use NIM~\cite{NIM} electronic modules (Discriminator, Timer and a Counter) to obtain the pulse rate of these also called Dark Counts. The minimum threshold of the discriminator is $\SI{-25}{\milli\volt}$. The resulting rates for the different samples are shown in \autoref{fig:SpuriousPulseRatesW4} and \autoref{fig:SpuriousPulseRatesW6}, for the Low and High Carbonated samples respectively. We can observe in these plots the frequency of the spurious pulses versus the bias voltage and in doted lines the operating voltage, that is the bias voltage needed to obtain a time resolution below $\SI{50}{\pico\second}$ (ATLAS requirements), being: $\SI{120}{\volt}, \SI{380}{\volt}$ and $\SI{540}{\volt}$  respectively for the fresh, $\fluenceUnits{0.8e15}$ and $\fluenceUnits{1.5e15}$ for the low carbonated samples and $\SI{120}{\volt}, \SI{390}{\volt}$ and $\SI{540}{\volt}$ for the high carbonated devices. Since we did not operate the most irradiated samples at higher voltages, we cannot determine an operating voltage for this yield point for either device. The frequency of spurious pulses in low carbonated samples is higher than this values, except for one of the two devices irradiated at \fluenceUnits{0.8e15}, and for the high carbonated spurious pulses appear earlier than the operative voltage in one of the devices irradiated at \fluenceUnits{1.5e15}.

\section{Runs comparison}

As mentioned above, this run\#15973 is the second production of CNM with carbonated devices, the first run with carbon enrichment in the gain layer was run\#15246, from which a complete characterization campaign was dedicated~\cite{CommonRunPaper}. In this section a comparison between these two runs is made. Technologically, the two main differences between these two runs are the Inter-Pad (IP) and the Junction Termination Extension (JTE). 

The IP is the geometric separation between adjacent zones in which the gain layer is implanted, thus being a no-gain zone and delimiting the sensor pixels. The IP used was \SI{57}{\micro\meter} and \SI{47}{\micro\meter} for run\#15246 and run\#15973 respectively. 

The JTE is a structure located at the edges of the gain layer, implemented to increase high voltage stability and improve gain homogeneity\cite{Curras}. A wider JTE improves the stability of the rupture as we approach the avalanche (so a smaller value increases instability). This phenomenon will be more pronounced at higher irradiation, as the applied voltage will have to increase if we want to maintain the gain, increasing the electric field at the periphery of the LGAD, a field that will be more difficult to control with a narrow JTE. In run\#15246 the JTE has a width of 15 microns and overlaps the multiplication 3 microns and for run\#15973 the JTE has a width of 10 microns and overlaps the multiplication 3 microns.

\begin{figure}
     \centering
     \begin{subfigure}[b]{0.8\textwidth}
         \centering
         \includegraphics[width=\textwidth]{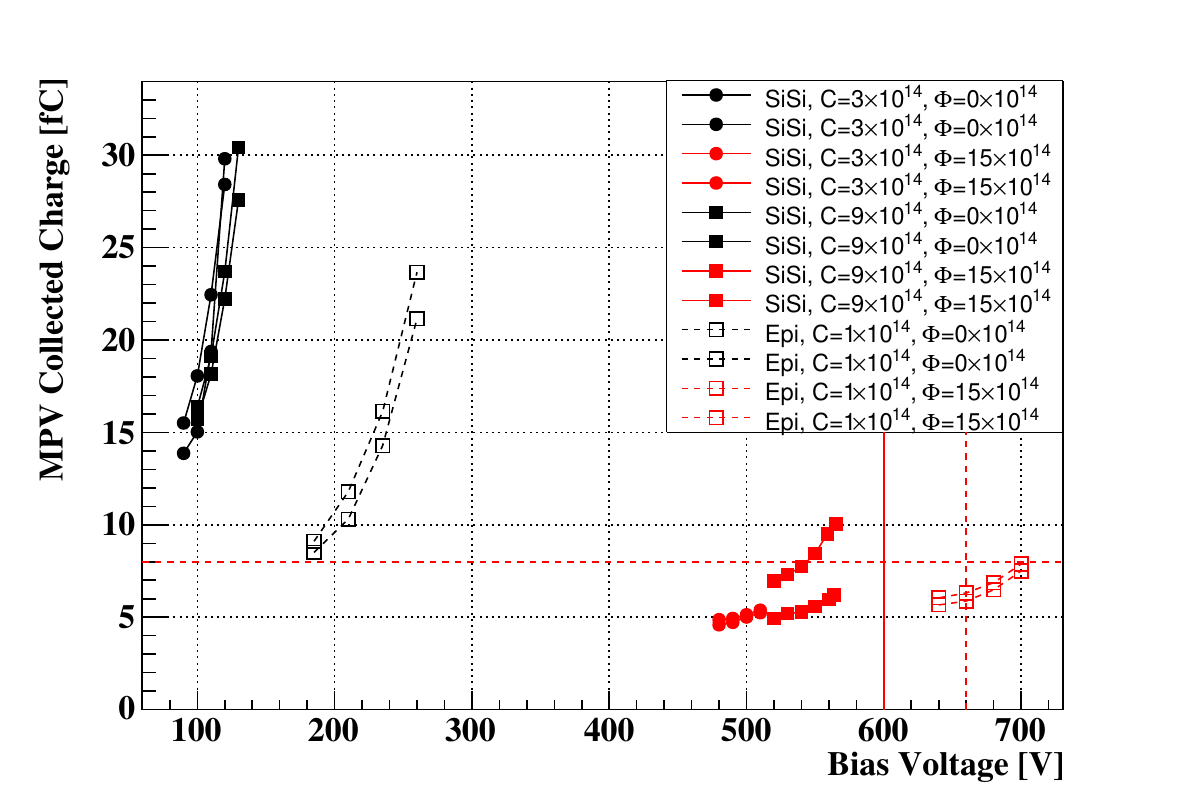}
     \end{subfigure}
    \caption{Collected charge of the devices from run \#15973 (filled markers) and run \#15246 (empty markers), irradiated (red) and non-irradiated (black) as a function of the bias voltage. Red lines are the limits for a bias voltage less than \SI{12}{\volt/\micro\meter} to prevent Single Event Burnout (SEB): vertical simple line is for run \#15973 and doted line for run \#15246, horizontal doted line indicates a time resolution of  \SI{50}{\pico\second} for both runs.}
    \label{fig:Charge_comp}
\end{figure}


\autoref{fig:Charge_comp} contains the samples from run \#15973 with different carbon doses are observed with filled markers. The carbonated devices from run \#15246 are represented by empty markers with circles for the low carbonated samples and squares for the high carbonated samples. It can be observed that the samples from run \#15973 exhibit a higher gain, as they generally require less voltage than the samples from run \#15246. It should be noted that in the case of irradiated devices, there is only one common fluence point (\fluenceUnits{15e14}) to be compared, and that the carbon doses are different between the runs under consideration.

\begin{figure}
     \centering
     \begin{subfigure}[b]{0.8\textwidth}
         \centering
        \includegraphics[width=\textwidth]{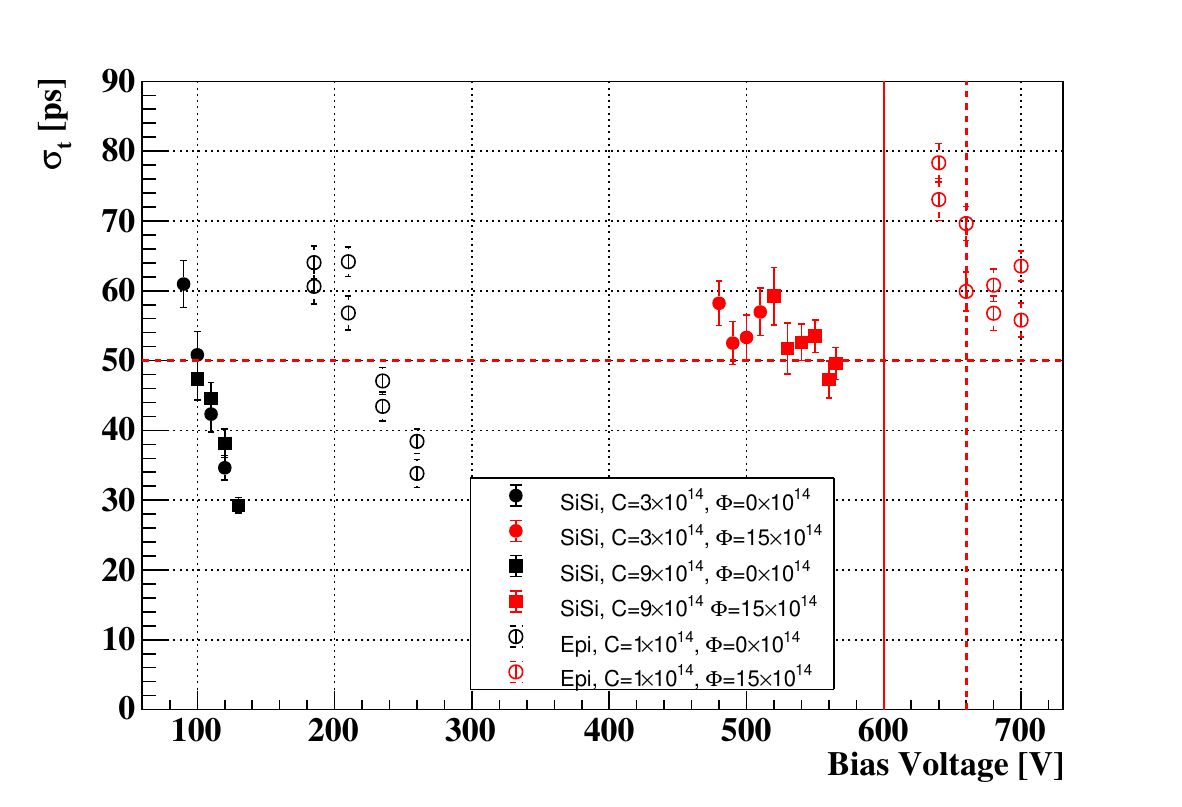}
     \end{subfigure}
    \caption{Time resolution of the devices from run \#15973 (filled markers) and run \#15246 (empty markers), irradiated (red) and non-irradiated (black) as a function of the bias voltage. Red lines are the limits for a bias voltage less than \SI{12}{\volt/\micro\meter} to prevent Single Event Burnout (SEB): vertical simple line is for run \#15973 and doted line for run \#15246, horizontal doted line indicates a time resolution of  \SI{50}{\pico\second} for both runs.}
    \label{fig:resol_comp}
\end{figure}

\autoref{fig:resol_comp} have the time resolution obtained from both runs with a narrow gain layer, irradiated and non-irradiated. In this case, the filled markers represent samples from run \#15973, fabricated with SiSi wafers, with circles indicating the carbonated samples with a dose of \SI{3e14}{at /\centi\meter\squared} and squares indicating the carbonated samples with dose of \SI{9e14}{at /\centi\meter\squared}. The empty markers represent the carbonated sensors from run \#15246, manufactured with epitaxial wafers. The sensors prior to irradiation from SiSi wafers (run  \#15973) exhibited a comparable operating voltage (independently of their carbon dose) and demonstrated the capacity to reach \SI{50}{\pico\second} at a bias voltage approximating \SI{100}{\volt}, a value that is smaller than the \SI{200}{\volt} necessary for the epitaxial sensors. The samples from run \#15973 (SiSi) irradiated at \fluenceUnits{15e14} exhibit superior time resolution in comparison to the preceding carbonated run \#15246 (epitaxial), although it should be noted that these sensors have higher carbon doses.

\section{Conclusions}
\label{sec:Conclusions}

In this study, the second manufacturing run at IMB-CNM of Low Gain Avalanche Detectors with a carbon-enriched multiplication layer were investigated for their radiation tolerance compared to conventional LGADs. The sensor were subjected to neutron irradiation at the TRIGRA reactor in Ljubljana, reaching a fluence of \fluenceUnits{2.5e15}. The results, reported in terms of degradation in timing performance and charge collection with increasing fluence, demonstrated the potential benefits of carbon enrichment in mitigating radiation damage effects, particularly the acceptor removal mechanism. The acceptor removal constant of carbonated samples with respect to the standard samples was reduced by more than a factor of two.

Time resolution and the collected charge was studied on the Radioactive Source (RS) setup for samples non-irradiated and irradiated up to fluences of \fluenceUnits{1.5e15}. As expected, degradation of the time resolution and the collected charge due to the irradiation was evidenced. The time resolution of the low Carbonated samples, at a fluence of \fluenceUnits{1.0e15} at the bias voltage of \SI{500}{\volt} achieved before the breakdown regime, is of \SI{52}{\pico\second} while  for the high carbonated LGADs is about \SI{47}{\pico\second} at a bias voltage of \SI{560}{\volt}. Confirming the better radiation tolerance of the high carbonated samples as it was the case of the acceptor removal coefficient.

Additionally, a  noise analysis was conducted on the samples. The investigation focused on the occurrence and frequency of micro-discharges, which may manifest as spurious pulses in silicon detectors due to thermal generation. The noise of carbonated samples was analyzed using a random trigger, measuring signal width without a radioactive source. The resulting noise values were examined across various fluences, as depicted in Figure \autoref{fig:SpuriousPulseRatesW4} and \autoref{fig:SpuriousPulseRatesW6}. Despite a more pronounced increase in noise for samples irradiated to \fluenceUnits{1.5e15}, the elevated noise levels did not impede the device's operation. Additionally, spurious, thermally generated pulses were measured beyond the operational voltages, showing a scaling trend with bias voltage for non-irradiated and low fluence samples, while higher fluence samples exhibited a quasi-constant region in the frequency. 

\section*{Acknowledgments}

This work was developed in the framework of the CERN DRD3 collaboration and has been funded by the Spanish Ministry of Science and Innovation (MICIU/AEI/10.13039/501100011033/) and by the European Union’s ERDF program “A way of making Europe”. Grant references: CEX2023-001397-M, PID2023-148418NB-C41, PID2023-148418NB-C42, PDC2023-145925-C31 and PDC2023-145925-C32. Also, it was supported by the European Union’s Horizon 2020 Research and Innovation funding program (Grant Agreement No. 101004761, AIDAInnova). 

This work has also been supported by the European Union NextGenerationEU/PRTR project C17.I02.P02 - SGI\_GICS\_Nuevas actuaciones en grandes infraestructuras de investigación europeas e internacionales, the Complementary Plan in Astrophysics and High-Energy Physics (CA25944),  project C17.I02.P02.S01.S03 CSIC CERN, funded by the Next Generation EU funds, RRF and PRTR funds, and the Autonomous Community of Cantabria.

This work has been developed in the framework of the Grant RyC-2023-044327-I funded by the Spanish Ministry of Science, Innovation and Universities (MICIU/AEI/10.13039/501100011033) and by FSE+, and co-funded by the European Social Fund program ‘‘El FSE invierte en tu futuro’’ with grant reference: PRE2019-
087514.


\end{document}